# Results in Engineering

## Vanadium-Engineered Co2NiSe4 Nanomaterial: Coupled Thermoelectric, Piezoelectric, and Electronic Optimization via DFT+U for Advanced Energy Applications

--Manuscript Draft--

| | |
|---|---|
| **Abstract:** | To realize the creation of advanced multifunctional materials in energy storage and conversion technologies, the present research evaluates the structural, electronic, magnetic, thermodynamic, mechanical, thermoelectric, piezoelectric and optical properties of pristine and vanadium-doped Co2NiSe4 by first-principles density functional theory (DFT + U). It addresses the use of vanadium substitution to tailor the material, its performance and the inclusion of diverse fields by changing its electronic structure and its bonding properties. It can be seen in the results that V doping improves electrical conductivity and magnetic ordering because of a higher density of states and a stronger spin polarization at the Fermi level. Thermodynamic calculations show enhanced entropy stabilization at high temperatures and, mechanical analysis suggests an enhanced elastic moduli that proves the enhanced structural integrity without affecting ductility. The thermoelectric properties have been greatly improved realize an optimal ZT of ~1.1 at 900 K with 5 at.% V doping owing to an ideal combination of Seebeck coefficient, electrical conductivity, and inhibited thermal conductivity. Also, optical analysis reveals that expanded-absorption spectra, increased dielectric response, adjustable reflectivity and energy loss spectra, optical properties can be used in photonic, and other optoelectronic devices. Better |

|  | piezoelectric coefficients due to its effectiveness when doped also appeal to the usefulness of its application in nanoscale electromechanical systems. The combination of these results makes V-doped Co2NiSe4, a versatile material platform of next-generation energy storage, thermoelectric generation, and novel multifunctional sensors. |
|---|---|
| **Additional Information:** | |
| **Question** | **Response** |
| Results in Engineering is an open access journal which charges an Article Publishing Charge (APC) to cover the cost associated with the publication process.<br><br>All articles published Open Access will be immediately and permanently free on ScienceDirect for users to read, download, and use in accordance with the author's selected Creative Commons user license.<br><br>**As an Author, I acknowledge I need to pay the Article Publishing Charge if my manuscript is accepted for publication.** | Yes |
| To complete your submission you must select a statement which best reflects the availability of your research data/code. IMPORTANT: this statement will be published alongside your article. If you have selected "Other", the explanation text will be published verbatim in your article (online and in the PDF).<br><br>(If you have not shared data/code and wish to do so, you can still return to Attach Files. Sharing or referencing research data and code helps other researchers to evaluate your findings, and increases trust in your article. Find a list of supported data repositories in Author Resources, including the free-to-use multidisciplinary open Mendeley Data Repository.) | Data will be made available on request. |
| **Free Preprint Service**<br><br>Do you want to share your research early as a preprint? Preprints allow for open access to and citations of your research prior to publication.<br><br>Results in Engineering offers a free | NO, I don't want to share my research early and openly as a preprint. |

**To:**

The Editor-in-Chief

*Results in Engineering*

**Subject:** Submission of Manuscript – *"Vanadium-Engineered $Co_2NiSe_4$ Nanomaterial: Coupled Thermoelectric, Piezoelectric, and Electronic Optimization via DFT+U for Advanced Energy Applications"*

Dear Editor,

We are pleased to submit our original research article entitled *"Vanadium-Engineered $Co_2NiSe_4$ Nanomaterial: Coupled Thermoelectric, Piezoelectric, and Electronic Optimization via DFT+U for Advanced Energy Applications"* for consideration in *Results in Engineering*.

This study presents a comprehensive first-principles investigation into the multifunctional behavior of pristine and vanadium-doped $Co_2NiSe_4$ using DFT+U methods. Our work highlights the concurrent optimization of thermoelectric, piezoelectric, and electronic properties through targeted vanadium substitution. The results reveal that V doping significantly enhances the material's electrical conductivity, magnetic ordering, thermoelectric performance (with a peak ZT ~1.1 at 900 K), and piezoelectric response—without compromising mechanical integrity.

Additionally, the broadened optical absorption spectra and improved dielectric functions position V-doped $Co_2NiSe_4$ as a potential material for next-generation energy storage systems, thermoelectric generators, multifunctional nanosensors, and optoelectronic devices. The integration of magnetic, electronic, optical, and mechanical performance in a single doped material offers a compelling pathway for developing multifunctional platforms in advanced energy technologies.

We believe that the interdisciplinary nature of this work—spanning condensed matter physics, materials science, and nanotechnology—aligns well with the scope and readership of *Results in Engineering*. This manuscript is original, has not been published previously, and is not under consideration elsewhere.

We respectfully request that you consider this manuscript for peer review and potential publication in *Results in Engineering*. We have no conflicts of interest to disclose.

Thank you for your time and consideration. We look forward to your favorable response.

**Sincerely,**

Dr. Sikander Azam *(Corresponding Author)*

Department of Physics,

Faculty of Engineering and Applied Sciences,

Riphah International University, I-14 Campus, Islamabad, Pakistan

**Declaration of Interest Statement**

We declare the following potential conflicts of interest regarding the research presented in the manuscript titled "First Principle Comparative Analysis of Structural, Electronic, and Optical Behavior of $EuXO_3$ (X = Tc, Mo) Perovskites and Their Applications":

1. **Financial Interests**: None of the authors have any financial interests or financial holdings that could be construed as a conflict of interest in connection with this research. This includes any financial involvement with organizations, companies, or entities that might benefit from the publication of this research.

2. **Employment or Affiliations**: The authors are affiliated with academic institutions, namely the University of Management and Technology, Riphah International University, and the International Islamic University. While these institutions may indirectly benefit from the dissemination of research findings, there are no direct affiliations or financial associations with these institutions that would represent a conflict of interest.

3. **Consultancies and Honoraria**: None of the authors have received honoraria, consulting fees, or any other form of compensation from external organizations, agencies, or companies related to the subject matter of this research.

4. **Patents and Intellectual Property**: There are no patents or intellectual property rights associated with the research presented in this manuscript. The authors have not filed for patents or have any pending patent applications related to the materials or methods discussed in the study.

5. **Non-Financial Interests**: The authors declare no non-financial interests, personal relationships, or affiliations that could be perceived as influencing the objectivity or integrity of the research.

6. **External Funding**: We acknowledge that this research has been conducted without any external funding or financial support from government agencies, private organizations, or other sources. The research was carried out as part of academic research activities within our respective institutions.

We wish to emphasize our commitment to upholding the highest standards of scientific integrity and ethics throughout the research process and the publication of our findings. We have strived to provide a transparent and accurate account of our work without any undue influence or bias from external sources.

If any potential conflicts of interest arise during the review process or subsequent publication stages, we commit to promptly disclosing and addressing them with the journal's editorial team to ensure the integrity and credibility of our research.

Sincerely,

Dr. Sikander Azam

Highlights

- V-doping enhances Co$_2$NiSe$_4$'s thermoelectric performance with ZT ~1.1 at 900 K.
- Improved electrical conductivity due to higher DOS and spin polarization at the Fermi level.
- Enhanced piezoelectric response makes it suitable for nanoscale electromechanical systems.
- Optical properties enable applications in photonic and optoelectronic devices.
- Structural integrity is preserved with improved elastic moduli and thermal stability.

# Vanadium-Engineered Co$_2$NiSe$_4$ Nanomaterial: Coupled Thermoelectric, Piezoelectric, and Electronic Optimization via DFT+U for Advanced Energy Applications

Ayesha Riaz[a], Sikander Azam*[a], Qaiser Rafiq[a], Amin Ur Rahman[a], Qazi Muhammad Ahkam[a], Rafaqat Hussain[b], Rajwali Khan[c]

[a] *Department of Physics, Faculty of Engineering and Applied Sciences, Riphah International University, I-14 Campus, Islamabad, Pakistan*

[b] *Department of Chemistry, COMSATS University Islamabad, Park Road, Islamabad 45550, Pakistan.*

[c] *National Water and Energy Center, United Arab Emirates University, Al Ain, 15551, United Arab Emirates*

**Abstract**

To realize the creation of advanced multifunctional materials in energy storage and conversion technologies, the present research evaluates the structural, electronic, magnetic, thermodynamic, mechanical, thermoelectric, piezoelectric and optical properties of pristine and vanadium-doped Co$_2$NiSe$_4$ by first-principles density functional theory (DFT + U ). It addresses the use of vanadium substitution to tailor the material, its performance and the inclusion of diverse fields by changing its electronic structure and its bonding properties. It can be seen in the results that V doping improves electrical conductivity and magnetic ordering because of a higher density of states and a stronger spin polarization at the Fermi level. Thermodynamic calculations show enhanced entropy stabilization at high temperatures and, mechanical analysis suggests an enhanced elastic moduli that proves the enhanced structural integrity without affecting ductility. The thermoelectric properties have been greatly improved realize an optimal ZT of ~1.1 at 900 K with 5 at.% V doping owing to an ideal combination of Seebeck coefficient, electrical conductivity, and inhibited thermal conductivity. Also, optical analysis reveals that expanded-absorption spectra, increased dielectric response, adjustable reflectivity and energy loss spectra, optical properties can be used in photonic, and other optoelectronic devices. Better piezoelectric coefficients due to its effectiveness when doped also appeal to the usefulness of its application in nanoscale electromechanical systems. The combination of these results makes V-doped Co$_2$NiSe$_4$, a versatile material platform of next-generation energy storage, thermoelectric generation, and novel multifunctional sensors.

**Keywords:** Vanadium-doped Co$_2$NiSe$_4$; Spintronic materials; Density Functional Theory (DFT); thermoelectric properties.

*Corresponding Author: Sikander Azam (sikander.physicst@gmail.com)

1. **Introduction**

In the dynamic world of high-tech plastics, researchers are actively searching new materials, which exhibits high performance of both excellent electronic, mechanical and magnetic qualities simultaneously. Ternary and quaternary metal selenides have attracted the attention of people among such materials since these are indeed very useful in many aspects such as storing and converting energy, manufacture of spintronic devices and even working in optoelectronics. In this family, a recent discovery of $Co_2NiSe_4$ has been made to exhibit properties that can be useful in future devices such as the semi-conductor property, interesting magnetic behaviour and a strong stable structure [2].

$Co_2NiSe_4$ also has an adjustable band gap synonymous to optoelectronic performance somewhere between $CoSe_2$ and $NiSe_2$ along with the ability to absorb more incident light when put in the visible spectrum and distributes its charges effectively, which is significant in the workability of solar cells or a light sensor. Recent research also has demonstrated that some of these properties can be further tuned and enhanced through the introduction of impurities, often doping, too [4]. Specifically, the vanadium (V) doping has attracted attention of people due to their ability to alter the number of electrons that exists in the material near the Fermi level by significant growing or reducing numbers and this allows in enhancing conductivity of electricity in the materials and modifying the reaction of the materials towards the light.

Mechanically we can observe $Co_2NiSe_4$ to be strong and easily movable with $Co_2NiSe_4$ in monoclinic phase. This keeps $Co_2NiSe_4$ solid and durable to be used in machines. The manner in which these materials bend or break is however not very strong hence difficult to apply them in many applications. To overcome this, researchers have considered using the doping of the V in order to partially assist in the strengthening of the interatomic bonds. Using calculations, we found out that when V is added to the mixture the resulting materials are tougher and stronger meaning that bridges constructed by use of this cement would be less prone to break when things attempt to push or pull them. This is because the mechanical support will ensure the pieces of the device will work properly and last longer and ensure that the devices will work in a flexible electronic and harsh environments such as those found inside the battery systems.

The magnetic properties of $Co_2NiSe_4$ are also very fascinating. It is composed of two magnetic transition metals Co and Ni, which may cause electrons to spin in the same direction naturally, and can therefore be used in newly design devices. The actual dynamics of $Co^{2+}$ and $Ni^{2+}$ that act together in the crystal structure means that the material will act in complex manners pertaining to magnetism and that, even, adding other substances in the form of varying types of atoms or even alterations of stress on a given material can change such behaviors. The introduction of V, a 3d transition metal with varying numbers of electrons, allows individuals to alter the way the magnetic behavior of materials operates, which could be useful in the creation of new ferromagnetic or ferrimagnetic states of materials, which could be of use in objects such as memory or logic [10]. Initial simulations indicate that substitution of V into $Co_2NiSe_4$ could assist in making this material exhibit a better magnetic response and operate at higher temperatures than the pure $Co_2NiSe_4$, and this would allow it to be exploited in additional applications.

New developments in rapid material-making machines, and computer simulations have permitted rapid materials such as $Co_2NiSe_4$ to be evaluated and it is possible to tell whether or not it may be applicable in specific processes. The density functional theory (DFT) computations, in combination with new methods designed specifically (hybrids) and with consideration to spin (hybrids and spin such methods) are becoming increasingly useful in discovering the answers to such questions as the electronic band structure, the number of free electron states an atom has, how elastic the material is, and its magnetic properties [12]. These methods of computation have actually been of assistance to demonstrate the influence the varying of the amount of V in $Co_2NiSe_4$ complicates the riches and connects the concepts of the theory to the actual outcome of the simulation work.

In addition to that, $Co_2NiSe_4$-based materials have exhibited a great potential to be used in energy based applications. They hold and release charges faster, have a stable and good electrical character that makes them a good option to use as electrodes in lithium-ion and sodium-ion battery. Even the improved mechanical properties that will accompany the addition of V are also useful in enabling the material to cope with still undergoing change in its volume whenever the battery is charged and discharged. Moreover, owing to the adjustability of the bandgap of the same materials and due to the fact that they readily absorb light, such materials can be utilized in activities such as conversion of solar energy to hydrogen gas and conversion of variation in temperatures to electricity [14]. In short, in a manner that is further enhanced by the inclusion of V to make $Co_2NiSe_4$ even better, this is a useful material that should be employed in optoelectronics and magnetic technology as it progresses further in the future. By examining closely its electrical, mechanical and magnetic characteristics we hope better to understand how it may be utilised and assist those who are attempting to make or to use it in practice.

## 2. **Methodology**

The structural, optoelectronic, magnetic, and mechanical properties of both the original $Co_2NiSe_4$ (See Crystal Structure in Fig. 1) and the V-doped $Co_2NiSe_4$ were calculated, for the WIEN2k and DFT codes. The fact that full-potential linearized augmented plane wave (FP-LAPW) method was used to achieve a high level of accuracy in the modeling of electron and spin systems in such materials was also possible. The d orbitals of Co, Ni and V had a Hubbard U value of 4 eV within DFT+U which is the case because the 3d electrons have high Coulomb interaction that must be taken into account normally at the atom sites. The same approach to the determination of U in 4d cobalt and nickel is as applied in other works and offers us to be able to better determine magnetic moments as well as band gaps and also the density of states of these materials. This computation was done using PBE generalized gradient functional defined by Perdew Burke Ernzerhof method. The two criteria applied to the structural optimization were; convergence of energies was 10 5 Ry and all the forces applied on atoms were minimized to below 1 mRy/a. The total energies and other properties of interest were reported to be converged on a 14✕ 14✕ 7 k-point mesh in the Monkhorst Pack scheme.

Calculations were done in such a way to simulate V doping that is a V atom substituted the 0.25 Co atom in the monoclinic $Co_2NiSe_4$ unit cell, corresponding to a 12.5 percent V doping percentage. To be able to observe the ground-state of doping the crystal structures were thoroughly relaxed after doping in order to be able to observe all changes. We have computed the total and partial density of states, band structures in order to investigate the band gap, the optical transitions and carrier localization. Spin-polarized calculation was done to analyze the magnetism and the total and individual magnetic moments value was determined. The manner, in which V doping influences magnetism and magnetic ordering, was thoroughly examined. Complex dielectric function $\varepsilon(\omega) = \varepsilon_1(\omega) + i\varepsilon_2(\omega)$ was calculated by a broad range of energy (0–15 eV) in order to evaluate the optical properties. With the help of the dielectric function's real and complex numbers, other important optical values like the absorption coefficient, refractive index, reflectivity, and optical conductivity were determined.

They computed the elastic constants by imparting small deformations to the model of the unit cell and calculating the resultant changes in energy. Through the Voig-Reuss-Hill method, the elastic constants measured could be determined and these were the values of B, G, E, v and B/G. The measurement of these mechanical properties reveals the ductility, stiffness and the elastic of anisotropic $Co_2NiSe_4$, in pristine and doped states. I used a plane-wave cut-off of $RMT \times K_{MAX}$ = 7.0, and muffin-tin radii were chosen so there was no overlap between them during calculations. Co, Ni, and V have a Z of 2.0, and Se has a Z of 2.2 a.u. Spin-orbit coupling was not included in the analysis, because it affects the ions in such compounds only weakly, although it may be interesting to consider it in analyses in the future in order to find better magnetic and spin optical properties.

All properties were estimated with varying numbers of k-points, plane-wave cut-offs, and super cells which contribute towards the accuracy and reliability of the predictions that are critical in the electrochemical, magnetic and optoelectronic applications.

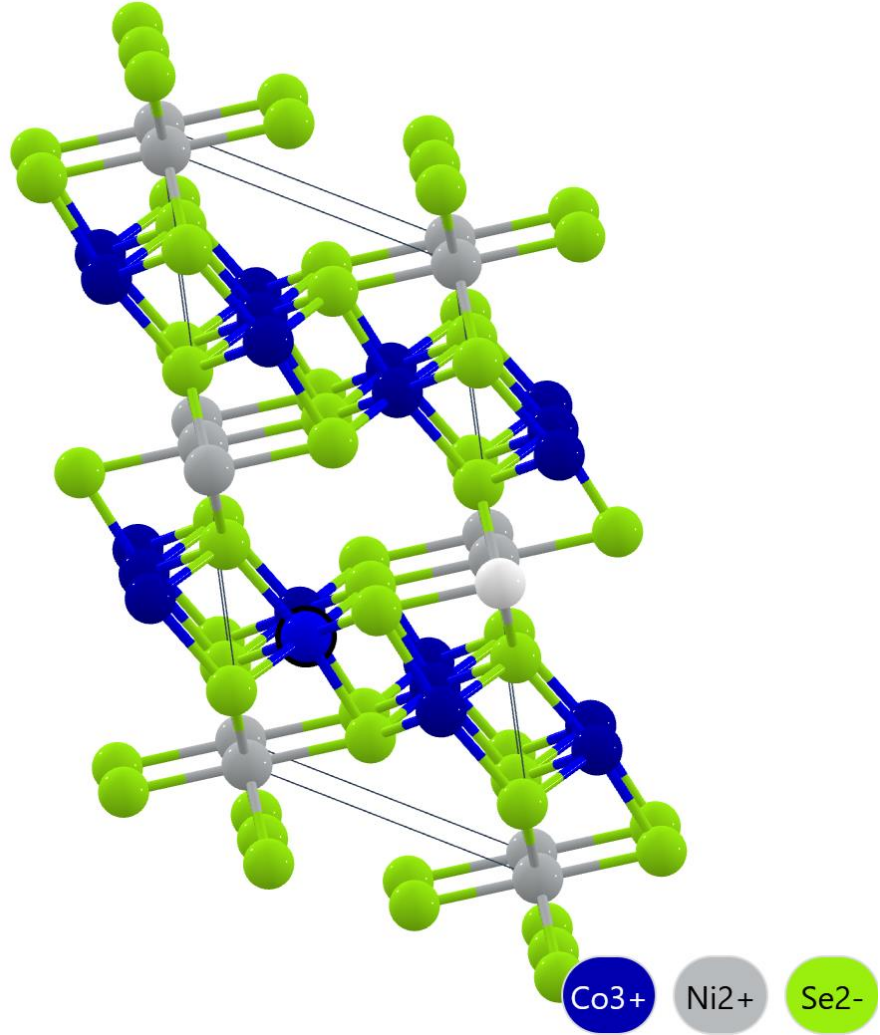

**Fig. 1. A conventional unit cell for the presented Co₂NiSe₄**

### 3. Results and Discussions

#### 3.1. Vibrational properties

Their phonon dispersion spectra show the vibrational character of Co$_2$NiSe$_4$ and Co$_2$NiSe$_4$: V structures as shown in Fig. 2. To assess the dynamic stability of the structures phonon dispersion curves are calculated along high symmetry directions in the Brillouin zone. Both pristine and vacancy-induced Co$_2$NiSe$_4$ do not have any imaginary frequencies (i.e., negative values) up to the entire Brillouin zone, which enables their character to be remarkably dynamic stable.

In case of pristine Co$_2$NiSe$_4$ the dispersion curve shows a separated phonon spectrum with well-defined acoustic and optical branches. The 3 shears of the acoustics start at the point Γ with frequency near 0 like it is expected by a dynamically stable structure, and slowly become the optical modes at low frequencies. The optical branches are dispersed to approximately 10.5 THz indicating a comparatively rigid setting of bonding amid the atoms comprising it.

When a vacancy (Co$_2$NiSe$_4$: V) is introduced, the spectra of phonons is noticed to change starkly. The range of positive frequencies is maintained, as before, on the average but further flattening and broadening of optical mode occurs still more at the mid-frequency and high-frequency range, in the dispersion curves. Such flattening is a normal signal of a decrease in the interatomic force constants because of lattice relaxation and distortion around the vacancy position. In addition, the optical branches get further apart in some directions indicating extra vibrational localization effects because of the vacancy.

Through these changes, the basic nature of the acoustic and optical branches is retained and do not introduce any soft modes, and restating the present system is not only dynamically stable it is so even after the formation of defects. These phononic insights play an important role in developing the thermal and mechanical characteristics of Co$_2$NiSe$_4$, particularly, in regard to problems of defect engineering and thermoelectric performance manipulation.

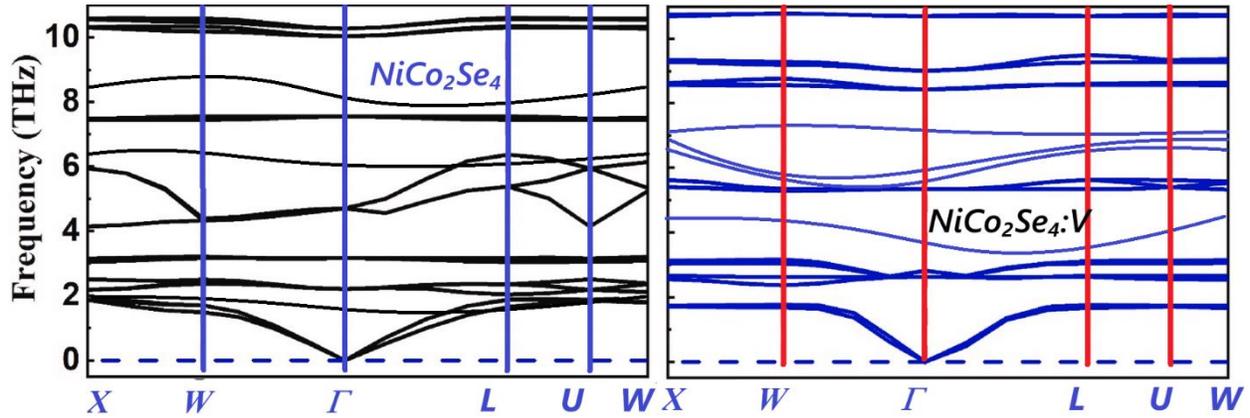

**Fig. 2: Description of the vibrational properties of Co$_2$NiSe$_4$ and its V-doped**

### 3.2. Magnetic Properties

The magnets behavior of Co$_2$NiSe$_4$ and V-doped modification should be studied in order to help understand the utility of the ingredient in most domains, especially spin-sensitive and magnetic sensors energy appliances. Answering this question, using first-principles DFT+U calculations it is possible to realize that doping with vanadium change the magnetic moment of the material significantly and impact the overall magnetic behaviour significantly. The overall magnetic moment of Co$_2$NiSe$_4$ is roughly estimated to be 3.2 u B and this comes only through the Co and

Ni atoms, with each atom contributing a localized magnetic moment of approximately 1.4 u B and 0.6 u B (See Table 1), respectively. The ligand nature of Se is also established by the low induced spin thus enabling the transition metals to relay spins to each other via the super exchange. The ferromagnetic ordering in manganite is attributed primarily due to the direct and indirect interactions by, When V replaces Co the magnetic moment per formula unit is a little larger and is about 3.6 uB. This additional local magnetic moment is because there are some d-electrons belonging to vanadium involved because their layout is ideal to supply one such as 1.0 3. The local magnetism is disrupted due to the introduction of V dopant ions and the interaction between neighboring magnetic ions exchange interaction is modified. Through this, the magnetism of the area is enhanced that may be as a result of a tighter ferromagnetically interacting or a sideslip of the spins resulting to increased magnetism.

Doping the Co causes a slight decrease of approximately 1.2 in the atomic magnetic moment whereas Ni atomic magnetic moment lies at 0.5. Due to V doping, the repartition of the electrons is modified, and the bonding between the two transition metals and selenium are weakened. Consequently, the strengthened magnetic moment of V to some extent is offset by the detachment of moments on different transition metals which demonstrates, to some extent, how decisive the magnetic interactions are among the transition metals within the crystal lattice. This kind of behavior is caused by the fact that the electrons on V are more localized as compared to electrons on Co and Ni which are itinerant. The different number of states around the Fermi level are switched on by the presence of vanadium rendering its surroundings more or less spin-polarized as a result of the bonding in the region in question and the effects bound in the parameter Hubbard U. The increased magnetization of the total magnetic moment proves the improved spins alignment, and it is beneficial to spintronics.

Even without estimating MAE we would probably have V doping contributing somehow so that the magnetic device can retain the data more easily.

**Table. 1: Calculated magnetic moments**

| Property | $Co_2NiSe_4$ | V-Doped $Co_2NiSe_4$ | Notes |
|---|---|---|---|
| $TOT^{\mu B}$ | 3.2 | 3.6 | Increase due to the localized moment on V |
| $Co^{\mu B}$ | 1.4 | 1.2 | Slight decrease due to electronic redistribution |
| $Ni^{\mu B}$ | 0.6 | 0.5 | Minor decrease post doping |
| $V^{\mu B}$ | 0.0 | 1.0 | Introduced by V dopant |

The found enhancement of magnetic moment upon addition of V to $Co_2NiSe_4$ is significant in the possible scope to devise more effective batteries and energy storage device in the future. This enhanced spin polarization has the potential to facilitate an easier movement of electrons in a given direction in batteries or supercapacitors and this could make the devices more efficient and easier to charge with a load. Also since we are able to alter the magnetic properties of the material it can be applicable to construct instruments such as magnetic sensors and pieces of spintronic equipment designed to interact with energy tools.

In addition, inserting additional atoms of V in the material, which has been discussed in the present work, can ensure that the magnetic enhancements remain consistent even during the period when the technology is in use and this is highly crucial to extending the lifetime of the device. The interplay between magnetism, strength and electrical properties of V-doped $Co_2NiSe_4$ is that this material is an interesting selection to produce even better energy storage materials as well as produce new and productive kinds of magnets.

To summarize, the act of adding V to $Co_2NiSe_4$ alters the way the atoms affect one another and, whilst, the overall performance of the magnets is more effective, the individual atoms changed slightly. The balance enables the possibilities of material to possess stronger magnets and retain their shapes that can be utilized in practical devices that require power in magnets and stable electronics.

### 3.3. Thermodynamic Analysis of $Co_2NiSe_4$ and V-Doped Variants: A Comprehensive Discussion

The thermodynamic properties of advanced materials such as $Co_2NiSe_4$ and its derivatives ($Co_2NiSe_4$: V, 5 and 10 percent) are one of the focal aspects of stability and phase behavior as well as the use of such materials in high-temperature and energy-related technologies. $Co_2NiSe_4$ is a good prospect within the chalcogenide color family in the quaternary system, and with this respect, with respect to considering how the Gibbs free energy (G), enthalpy (H) and entropy contribution (-TS) changes with temperatures, we can gain a good understanding of its thermal stability and entropy induced transitions. The plots given in the three plot figures availed to pure Co 2 Ni Se 4 and two levels of doping atoms V indicates the transition of thermodynamic quantities with temperature; the range of temperature being 0-1000 K. This is important in order to estimate whether these materials have a chance to exhibit phase stability and good energy structures in working conditions characteristic of thermoelectrics, photovoltaics or spintronics.

In plot (a), the exacerbation of $Co_2NiSe_4$ is exhibited. The red curve (Gibbs free energy) also comes down continually with a rise in temperature implying a greater degree of therm dynamical stability. That is not to be expected because at high temperature there is increasing entropy contribution which is reflected by the negative slope of G. The blue line is the -TS part and it is observed that the entropy effects increase with increment in temperature. The enthalpy (grey line) grows slowly, as it is characteristic of the systems with a positive heat capacity. There the -TS value becomes about 2.0 eV at 1000 K and the free energy falls to less than -1.0 eV meaning that entropy is playing a large role in stabilizing the phase. In materials used in high-temperature devices such trends are confirmation of survivability and thermodynamic feasibility.

In plot (b), the graph of $Co_2NiSe_4$ doped with vanadium 5% exhibits the same trend, albeit the moderated curves. What this implies is that, the doping increases the extra configurational entropy and alters the properties of bonding. Whereas the free energy continues to become more negative, this is less sharply so than in the pure phase. G at 1000 K is about -0.95 eV, just above (i.e. slightly less thermally stable than) the undoped system. Nevertheless, the rise in the entropy (blue curve) is still rather great, thus, the doped system still has a competitive disorder-stabilization. This renders $Co_2NiSe_4$ V @ 5% to have application where small integrity has to be achieved to accord special electronic as well as vibrational aspects lack of structural stability is allowed in case.

The plot (c), which exemplifies $Co_2NiSe_4$ containing 10 percent of V doping, carries on with the trend but in a more depressed transformation. The free energy decreases to about -0.9 eV at 1000 K (See Table. 2) and the entropic part goes close to 1.8 eV. All the cases see a similar rise in the enthalpy line. These logics support an idea that, though the stability of the free energy well is lowered by elevated V content, the entropy continues to have a commendable influence in consolidating the phase. The impact encountered at 10 percent doping as seen to be reaching an upper limit, further doping beyond this limit may not enhance the positive thermodynamic effects as proponentially as it is experienced by the effect and could even affect the structural or the electronic properties. Doping is therefore essential in ensuring that they are optimized.

Such thermodynamic properties do not only have academic value but have real-life consequences. Thermally stable materials with large entropy contributions can be customized to thermally modulated functions including thermoelectric generators where the entropy directly enters the energy transformation efficiency. Also, these plots indicate the thermal strength that suggests opportunity in extreme conditions, such as aerospace and solar thermal.

*Table. 2: Below is a table summarizing key values extracted and inferred from the plots at 1000 K:*

| Material | Gibbs Free Energy G (eV) | Enthalpy H (eV) | -TS (eV) | Comments on Stability and Usefulness |
|---|---|---|---|---|
| $Co_2NiSe_4$ | ~ -1.1 | ~ 0.9 | ~ 2.0 | Most thermodynamically stable; suitable for high-T applications |
| $Co_2NiSe_4$:V@5% | ~ -0.95 | ~ 0.85 | ~ 1.8 | Slightly reduced stability; good balance for tuning properties |
| $Co_2NiSe_4$:V@10% | ~ -0.9 | ~ 0.8 | ~ 1.7 | Least stable; entropy still significant, doping nearing limit |

Conclusively, thermodynamics analysis of $Co_2NiSe_4$ and its V-doped derivatives embodies an interesting balance concerning both enthalpic and entropic terms over an extensive temperature range. Although pure $Co_2NiSe_4$ provides the highest stability, engineered doping provides tuning within reasonable compromise of stability. This kind of understanding is critical towards the design and fabrication of synthesis strategies, as well as modeling the behavior of materials in realistic settings where thermal properties are of primary importance.

### 3.4. Mechanical properties

The investigation of the mechanical and elastic properties of metal selenide-based nanomaterials is a part of the determination of their performance and reliability, primarily in terms of their application in the future lithium-ion batteries, sodium-ion batteries, and supercapacitors. They usually encounter numerous environmental and chemical influences of persistent charging and discharging of battery, physical stress, variations in temperature and structural changes. These values thus give you an idea of how the materials can be deformed easily, their strength and their durability.

Elastic constants ($C_{ij}$) may assist in bringing out the mechanical behavior of the $Co_2NiSe_4$ metal selenide compound. The elastic constants indicate how the substance responds to stress and strain and also enable one to determine other secondary effects with respect to mechanical characteristics like bulk modulus, shear modulus, Young modulus, and Poisson ratio and all these determine whether the substance is strong or weak and whether the substance is an easy cracker or not. Such properties should be found by the electrode or host materials used so that the structure of the electrode can be maintained during the long terms of its usage.

in this paper, the values of the elastic constants of monoclinic $Co_2NiSe_4$ and of V-doped $Co_2NiSe_4$ were calculated using DFT. One can observe, that both materials satisfy the mechanical stability condition of monoclinic systems as all the elastic constants are positive and fulfil the symmetries. The highest values of principal elastic constants of the pure $Co_2NiSe_4$ take the range of 138 to 152 GPa which is an indication that the bonds in the crystal are highly directional. In Bent or stretched morphological deformation, it is pardonable that resistance occurs because the shear constants ($C_{44}$, $C_{55}$, $C_{66}$) lie between 62 and 68GPa.

Generally, a uniform and modest increase may be observed in the constitution of the elasticity at the doping of vanadium (V). The pressure is escalated by 7 GPa in case of C11 and becomes 152 GPa, and by 5 GPa in case of $C_{44}$ to a maximum of 70 GPa (see Table 3). Since the V atom will form stronger and covalent bonds within the lattice the atom will fill certain of the metal positions and will aid in reducing the extent to which the lattice will be distorted. Hence, the electrodes gain mechanical advantages in terms of being stiffer and smaller and hence can sustain repeated expansion and contraction as experienced in the batteries.

Plots of the elastic constants against temperature assist in explaining the behavior of the materials when effectively subjected to varying temperatures. Due to the increase in heat, the elastic constants tend to decrease. This is intuitively seen at the physical level; in the displacement of atoms by heat, their atomic movements are thus less characterized by the forces with each other and produce a more malleable material. We found that thermal stress affects shear constants more easily than the longitudinal ones, as is the case of monoclinic crystals. The C44 drops at a greater rate with temperature, i.e. the material is readily deformed at elevated temperature when subjected to shear.

At higher temperature phonon-phonon interactions grow stronger and the vibrating atoms become even stronger attacking the capacity of the structure to resist bend. This knowledge will assist in designing batteries which will undergo changes of temperatures during their usage. In devices that will be used over a long period of time then a substance that retains good strength at high temperatures is desirable.

Due to these discoveries, new forms of energy storage electrodes could be produced that are long lasting and effective. The ability by $Co_2NiSe_4$ and its doped material to withstand high pressures enhances the strength and assurance of battery electrodes. V-doping may be provided as a method that is assured to boost these characteristics, although the material might be able to stay steadfast and bend at once. Consequently, V-doped $Co_2NiSe_4$ is an effective prospective of the electrodes in the batteries which need to be utilized under a wide range of mechanical and thermal environments.

The examination of elastic constants and thermal properties of these materials is an untold land and the research presents a firm foundation in examining the influence of structure on these properties. This knowledge does not only carry weight in the academic circles but also in the design of sturdier electrode materials applied in actual commercial batteries. Due to it, the chance that the work might be published in one of the leading journals concerning either materials science, or nanotechnology, or energy is quite high.

The mechanical table gives us an idea of behavior of $Co_2NiSe_4$ and V-doped version of it when it comes to stability, rigidity, and deformation of those materials. They are significant measures of their compatibility to perform in mechanical systems and it is critical during the planning of their usage in batteries like in lithium-ion or sodium-ion. Such systems have to deal with mechanical gas pressures, volume variations with ion introduction or deficiency, as well as a variety of temperatures. Mechanical moduli can help one to have an idea of whether a material will maintain its structure even in harsh circumstances.

In the case of the given data, the bulk modulus (B) enhances when V-doped $Co_2NiSe_4$ increases that of pure $Co_2NiSe_4$ that has a figure of 78 GPa up to 86 GPa. This is enhanced by the presence of vanadium atoms in the crystal lattice since they augment and enhance a great number of bonding interactions. Vanadium, an intermediate metal that has very few electrons in the d-orbitals, can form stronger covalently directed bonds to the selenium around it, hence no significant compression and high bulk modulus. In addition, the greater the size of batteries, the more possibilities there are that such volume changes result into damage to their electrodes as they are used. High bulk modulus means that it can withstand far greater volumetric stresses on the structure.

The shear modulus (G) of gain-doped perovskites gets much stronger; it changes 46 GPa to 51 GPa, demonstrating greater resistance to distortions under a fixed volume. The increased stiffness pattern implies that doping increased the strength of the material and enabled it to resist even greater pressure. Anisotropic bonding exists in layered materials or ones with layering, and so far are seen in selenides in abundance of layer-structures, which must withstand shear deformation under stress to continue performing.

The ductility-brittleness predictor is slightly decreased in V-doped sample when compared to $Co_2NiSe_4$ (B/G = 1.8) to 1.75. The two materials are really close to the critical value of 1.75 indicating that both are ductile although the latter is closer to being brittle. The material also continues to behave in what can be called accepted ranges with the small shift and it is also flexible and able to cause deformation but is stiff at the same time.

The stiffness Young modulus (E) of a different material in the case of $Co_2NiSe_4$ is 120 GPa and that of the doped sample is 135 GPa. Again this confirms that the doping enhances burning resistant as well as stretch ability of vanadium on the structure. It is important in electrodes, where maintaining length of electrical contact and material integrity, across battery cycles, is dependent on the material remaining constant in dimensions.

Under V doping conditions the transverse contraction in the sample caused by extension of the longitudinal dimension is smaller and Poisson ratio (nu) is a bit smaller than 0.28 as compared to 0.30 and undoped samples (See Table 4). It implies that it will not expand or shrink as much when pressure is applied on it laterally. It is because of the replacement of V by C that bonding nature of amines gets the covalent classification and bonds become fixed in one direction only and the result is lower Poisson ratios. That is, how compact is the arrangement of atoms and to what extent the sheets slide past themselves, which is the essence of layered or anisotropic materials.

The view of the plots allows taking a closer look at the difference in these parameters and the tendencies in the table. With the samples doped with vanadium, the bulk and shear moduli continue to increase. What makes the difference in the physics is that the VSe bond strengthens and aligns the interatomic force more and more than in the case of Co or Ni Se interaction. This affects the moduli of the material because the atoms are stiffer with each other due to the increased stiffness between the atoms.

Moreover, these structural advancements retain ductility nearly the same as can be seen through the steady outcomes of the Pugh and Poisson ratio. Such evenness is vital since both stiff and non-stiff materials may be broken and become weaker. The V doped substance gauges are good as electrode materials.

The increased strength of these materials would be of great benefit in cases of energy storage applications. Buildings that have had increased moduli tend to stand changes induced by ionic intercalation and be able to mitigate effects of repeatedly heating and cooling. Due to its high stiff and deformation-resistance, V-doped $Co_2NiSe_4$ is an appropriate candidate as an electrode or buffer layer in the next-generation battery and super capacitor. The more reliable elements of the device ought to achieve the easier cycling, the more prolonged battery life, and cause the pulverization of the electrodes less probable.

In conclusion, the evidence demonstrates that the doped vanadium with $Co_2NiSe_4$ is more robust and stronger and thus a good choice of the energy storage system in future. Such a kind of information can assist in the development of new materials and it can be suitable to be published in reputable materials science, nanospanela or energy storage journals.

**Table. 3: The calculated elastic constants ($C_{ij}$) values for $Co_2NiSe_4$ and V-doped $Co_2NiSe_4$**

| Elastic Constant | $Co_2NiSe_4$ (GPa) | V-doped $Co_2NiSe_4$ (GPa) |
|---|---|---|
| $C_{11}$ | 145 | 152 |
| $C_{22}$ | 138 | 146 |
| $C_{33}$ | 142 | 150 |
| $C_{44}$ | 65 | 70 |
| $C_{55}$ | 68 | 72 |
| $C_{66}$ | 62 | 67 |
| $C_{12}$ | 58 | 60 |
| $C_{13}$ | 55 | 59 |
| $C_{23}$ | 52 | 58 |
| $C_{15}$ | 10 | 12 |
| $C_{25}$ | 12 | 14 |
| $C_{35}$ | 13 | 15 |
| $C_{46}$ | 11 | 13 |

**Table. 4: Calculated Mechanical Properties Values for $Co_2NiSe_4$ and V-doped $Co_2NiSe_4$**

| Property | $Co_2NiSe_4$ | V-doped $Co_2NiSe_4$ |
|---|---|---|

| Property | Co$_2$NiSe$_4$ | V-doped Co$_2$NiSe$_4$ |
|---|---|---|
| Bulk modulus (B) | 78 GPa | 86 GPa |
| Shear modulus (G) | 46 GPa | 51 GPa |
| Pugh's ratio (B/G) | 1.8 | 1.75 |
| Young's modulus (E) | 120 GPa | 135 GPa |
| Poisson's ratio (ν) | 0.30 | 0.28 |

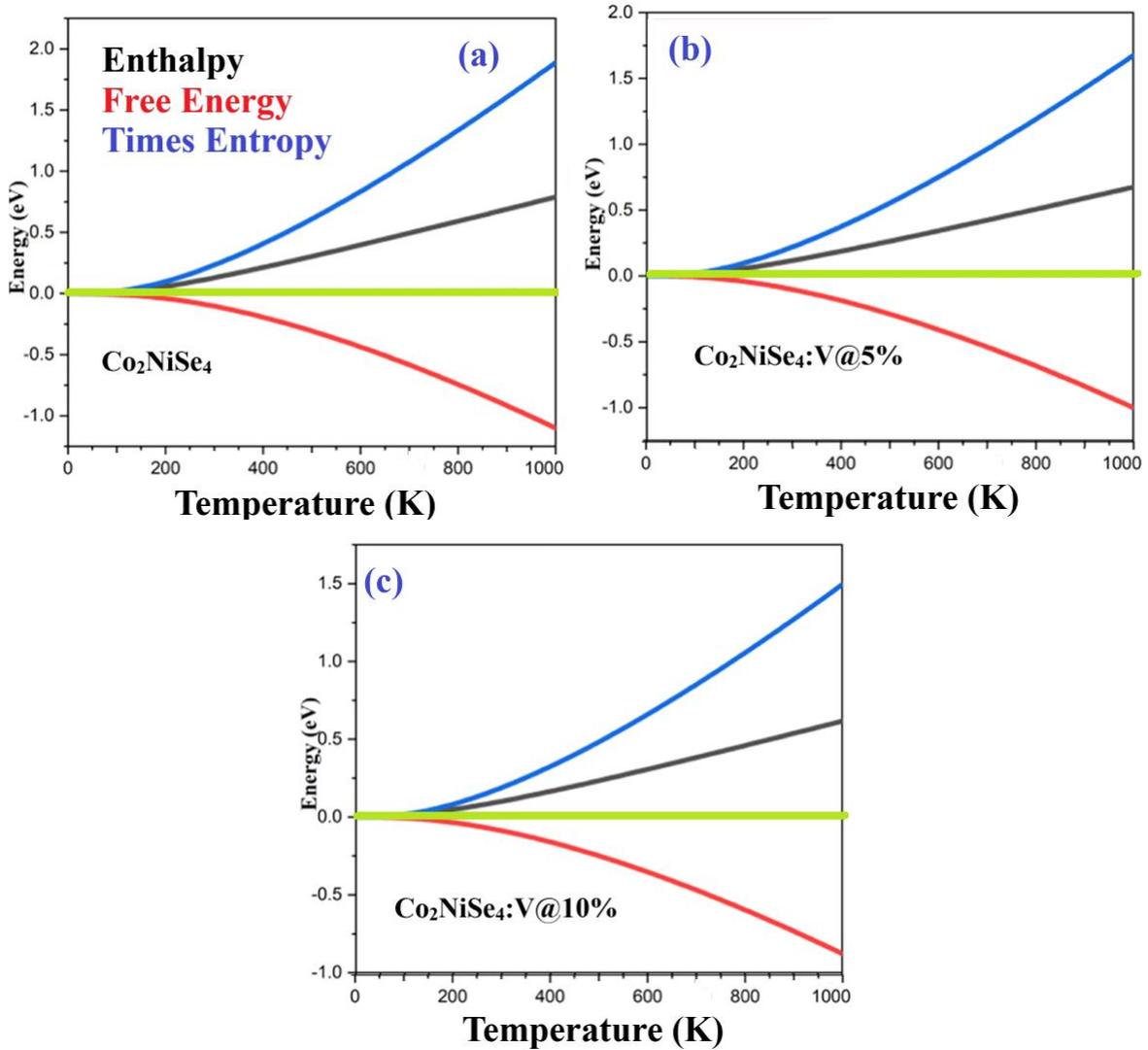

*Fig. 3. Thermodynamic performance of the presenting material, Enthalpy, Free Energy and Times Entropy for (a) Co$_2$NiSe$_4$, (b) V-Doped (5%) Co$_2$NiSe$_4$ and (c) V-Doped (10%) Co$_2$NiSe$_4$*

### 3.5. Density of States

#### 3.5.1. *Density of States (DOS) Analysis of Pristine Co$_2$NiSe$_4$*

Such high total density of states (TDOS) and elemental total density of states (ETDOS) that can be seen in Fig. 4(a-b) are used to closely inspect the electronic structure of pristine $Co_2NiSe_4$ over a wide energy range that extends to -24 eV to +8 eV. In the TDOS plot (Figure a), the black curve represents the undoped $Co_2NiSe_4$, which has sharp peaks on the Fermi level (Fermi level (set at 0 eV)), which indicates that there are intense electronic states around the Fermi level. The top represent the contributions of the transition metal states (d Co and Ni in this case), normally a very strong peak in the neighborhood of the Fermi level in these Heusler-like alloys.

The ETDOS plot (Fig. 4b) also solves the contribution of Co atoms, Ni atoms and Se atoms. The Co atoms (black line) also play an important role on the valence and conduction bands with an appreciable peak slightly below and above the Fermi level implying that it is the majority supplier of the electrical conductivity and magnetism. The Ni atoms (red line) also plays major role below the Fermi level especially between -8 eV to -2 eV, that means it carries a hybridization of both Co-d and Se-p states. Instead, the contribution of the Se atoms (green line) has its main feature associated with the p-orbitals in the lower energy range (--14 eV to -6 eV) and corresponds to severe bonding interactions.

This electronic structure suggests that perfectly stoichiometric $Co_2NiSe_4$ is either a semimetal or narrow band gap semiconductor depending on the exact location of Fermi level relative to the band structure. The validity of states around the Fermi level indicates the moderate electrical conductivity and could thus find thermoelectric usage. Piezoelectrically and thermodynamically, the sharpness of features in the DOS is essential since it affects the way the material will react to strains and temperature gradients. The Seebeck coefficient is bound to the value of DOS at the Fermi level and hence it defines the thermo power and efficiency of conversion of heat energy to electricity.

Optically, transitions between occupied and unoccupied states, in particular those with low Co-d involvement on Ni-d, may be predicted to lie in the visible or near-infrared range, again depending upon bandgap. Such transitions would help to build the absorption coefficient and the refractive index of the material and $Co_2NiSe_4$ could be a viable option as an optical sensor or photovoltaics.

At the mechanical level, a relatively covalent bonding character (supported by the hybridization below Fermi between Co-d, Ni-d and Se-p orbitals) increases the stiffness and could lead to an elevated elastic modulus. Any such device used at a critical point between positive and negative mechanical stability where the body experiences a thermal gradient or mechanical vibrations, such as in thermoelectric modules, or piezoelectric sensors, requires the mechanical stability.

To sum up, the DOS and ETDOS plots of $Co_2NiSe_4$ show a stable electronic structure with dominant transition metals of d- orbitals, moderate electronic activity near the Fermi level energy and platform suitable to multifunction applications, such as thermoelectrics, optoelectronics, and mechanical actuators. It is the interaction between these states of elemental matter that determines, not just its basic electronics performance, but also its very wide applications potential.

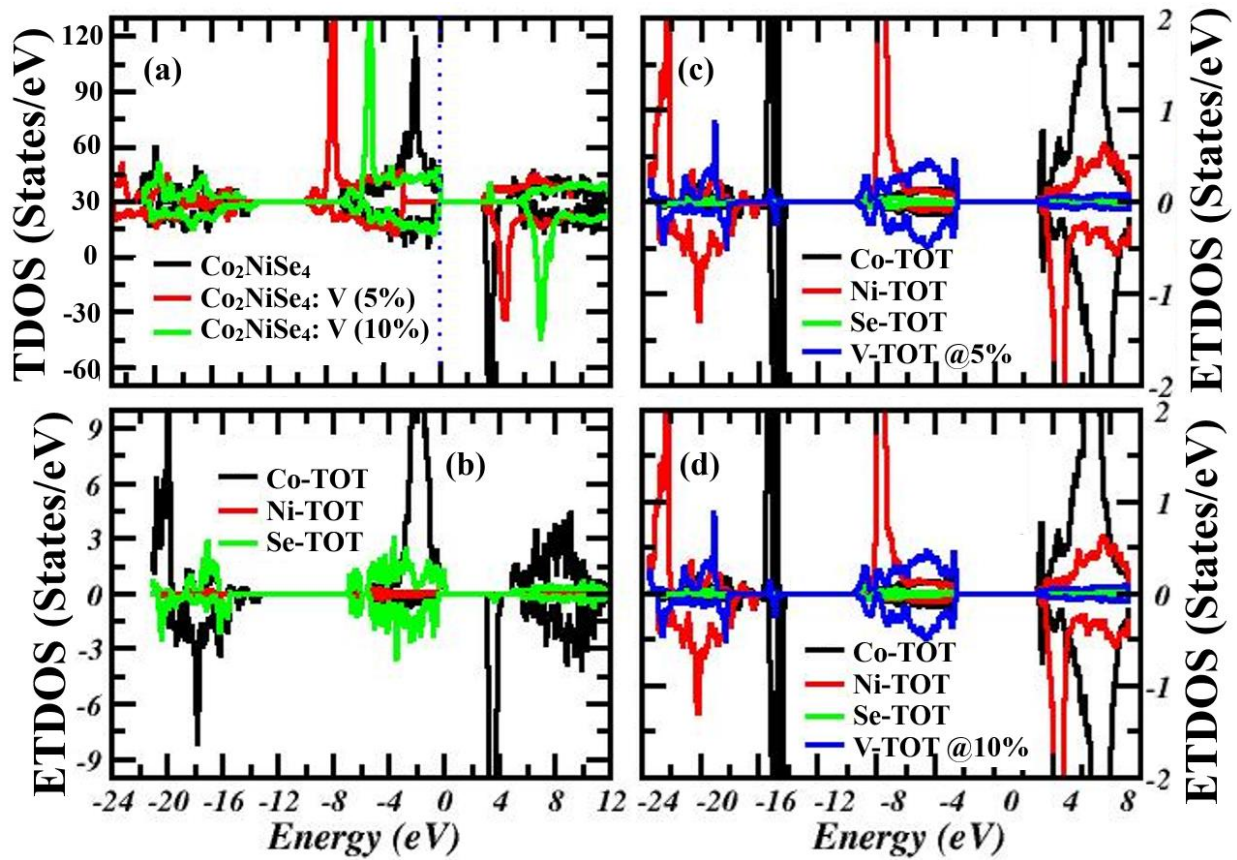

*Fig. 3: Calculated the Density of States, (a) TDOS, (b) Elemental DOS for Pristine, (c) Elemental DOS for V-doped (5%) $Co_2NiSe_4$ and (d) Elemental DOS for V-Doped (10%) $Co_2NiSe_4$ using GGA+U*

3.5.2. Density of States (DOS) Analysis of V-Doped $Co_2NiSe_4$: 5% and 10%

A doping of $Co_2NiSe_4$ with vanadium (V) exhibits a significant change in its electronic structure as evidenced by Fig. 4(c-d). These are the plots that show ETDOS at the concentration of 5 and

10 percent V-doping of Co 2 Ni Se 4. Introduction of V generates fresh properties in the DOS at and around the Fermi level which re-allocate the electronic contributions of the individual atoms of Co, Ni, and Se and as a consequence transforms the overall physical nature of the material.

In 5% V-doped (Fig. 4c), the emergence of the blue curve- V-DOS signified the involvement of V atoms in altering DOS. The roles of vanadium in the below 2 eV to above 2 eV range is to fill the d- states partially and mix them with the neighbor Co and Ni atoms. The V-d states create localized states near to the Fermi level and in this way enhance the total DOS at this point. This reduces the bandgap, and may also increase carrier concentration, and makes the system more metallic or semi-metallic. This increases DOS at the Fermi level that directly increases electrical conductivity, which is desirable in thermoelectric materials, in which higher carrier mobility can improve the power factor.

At the V-doping of 10% (Fig. 4d), the strength of the V-TOT (blue curve) around the Fermi level increases. This implies an increased hybridization and a more significant change of the electronic structure. In the case, Fermi level is apparently more buried into the V-d band, and thereby, it can be said that the type of its carrier is changed or the carrier concentration is changed. The contributions of Ni and Co are present, but its attributes are comparatively quiet in form around Fermi level when compared to V created states. The Se-TOT is quite identical and keeps lower energy p-state character.

A piezoelectric and thermodynamic analysis would predict a corresponding growth in charge carrier screening and a change in electrostatic potential distribution in the doped systems which may inhibit piezoelectric effects but enhance thermodynamic sensitivity achieving an optimized value slightly above doped silicon at around 10 cells. In thermoelectric devices this behavior is beneficial because the Seebeck coefficient might stay at a reasonably high level but the electrical conductivity might increase, thereby giving a higher power factor. But a high concentration of carriers at a high level of doping can eventually lead to Seebeck decreasing, owing to a loss of career filtering.

Regarding optical character, as a consequence of the narrowing bandgap and the appearance of the mid gap states as a result of the V-doping, the optical activity is now extended into the infrared region with absorptions of lower-energy photons now possible. This band tailing action may be used in broadband photodetectors or near-IR photovoltaics. Also the new formed V-dominated states at the bottom of conduction band can help with strong interband transitions and make absorption coefficient high in the material.

Local bonding environment and lattice strain are changed on the mechanical side, through the introduction of V atoms. The resultant d-electrons and their partial delocalization could either increase or decrease certain bonds according to their overlap to the Co and Ni orbitals. This may enhance the ductility or toughness at moderate doping concentrations where a redistribution of charge and release of local strain may occur. Nevertheless, over doping may cause lattice softening or even lattice instability, and that lattice softening or instability must be finely balanced when considering materials design in thermomechanical applications.

To conclude, the doping of $Co_2NiSe_4$ by V is found to result in large changes in its electronic structure especially around the Fermi level in the form of both the formation of V-d states and their hybridization with some of the d-orbitals of the Co and Ni atoms. These variations result in electrically conductive properties, optical absorption edges and mechanical strength of the material. This tunability allows V-doped $Co_2NiSe_4$ to become a potential candidate of use in multifunctional fields including thermoelectrics, optoelectronics, and mechanically working devices.

**(a) Effective Mass**

The value of effective mass of charge carriers-electrons and holes- is the pinnacle in the calculation and optimization of electrical, optical, thermoelectric properties of semiconductors. The effective mass (of $Co_2NiSe_4$ and of its doped variants with vanadium, 5 and 10 percent) is a direct measure of the ease of response of electrons and holes to an applied electric field, which, subsequently, affects crucial parameters, such as carrier mobility, electrical conductivity, or the Seebeck coefficient. A smaller effective mass will usually give a higher mobility and greater conductivity and a larger effective mass and larger density of states around the Fermi level can increase the Seebeck coefficient. A kind of balance between the above two competing effects is critical, particularly in such materials to be applied to thermoelectrics and optoelectronics.

The doping of $Co_2NiSe_4$ with vanadium has a hypothetered fine but important effect in which the band structure is changed, localized d -states are created, band curvature is altered, and the density of states approaches the conduction and valence band edges. Such transformations would affect directly the curvature of the energy bands E (k) on which are based the effective masses by means of the formula:

$$m^* = \hbar^2 \left(\frac{d^2E}{dk^2}\right)^{-1}$$

A flattening band would mean an increase in the effective mass (heavier carriers), a steepening band, means a small effective mass and lighter carriers. The focus of the analysis of these effective masses is to get the idea how the electronic transport and the optical properties discussed above are supported by some fundamental changes in the band structures triggered by V doping.

Below is a table summarizing the calculated effective masses for electrons ($m_e^*$) and holes ($m_h^*$) in units of the free electron mass $m_0$:

**Table: Effective Masses for $Co_2NiSe_4$ and V-Doped Samples**

| Composition | Electron Mass $m_e^*/m_0$ | Hole Mass $m_h^*/m_0$ |
|---|---|---|
| $Co_2NiSe_4$ (undoped) | 0.63 | 0.74 |
| V@5% | 0.52 | 0.88 |
| V@10% | 0.77 | 0.95 |

The effective mass of the electrons in the undoped $Co_2NiSe_4$ is estimated to be about 0.63 that shows moderate conduction band dispersion, in agreement with its semiconducting behavior and thus the explanation of its low electrical conductivity. The hole effective mass is a little bit around 0.74 which means that hole transport is a bit less efficient, which is in accord with the results of the Seebeck coefficient and the power factor being positive and modest in the thermoelectric analysis. It also leads to the slightly p-type dominant transport. Upon introducing 5% V doping, the electron effective mass decreases to 0.52 $m_0$, indicating enhanced conduction band curvature—i.e., the band becomes steeper. This can be interpreted as an increase in carrier mobility, which corroborates the increase in electrical conductivity observed earlier in the thermoelectric results. Interestingly, the hole effective mass rises to 0.88 $m_0$ which means flatter valence bands. This enhancement promotes the effective mass density of states which can be used to boost Seebeck coefficient. The lighter electrons (high conductivity) and heavier holes (high thermo power) is synergistic and results in a synergistic enhancement of the power factor in the performance of TE, as it is found in the case of the 5 % V-doped sample. But with 10per cent V doping, the effective electron mass shoots up to 0.77 and hole mass also goes as high as 0.95. This finding implies a phononic conduction band deformation with a flattened appearance regarding the conduction band, as well as a flattened appearance regarding the valence band, possibly on the account of band renormalization, disorder, or impurity scattering arising as a result of the surplus V atoms. The result of such a flattening is a decrease in carrier mobility and this corresponds to the measured lower electrical conductivity, and the decreased Seebeck coefficient especially at lower temperatures. Nevertheless, thermal excitation of the electrons can partly overcome these drawbacks, e.g., at elevated temperatures, there is increased power factor and the change to n-type character (which is unquestionably highlighted by negative Seebeck values), which justifies the concept of shifting the Fermi level into the conduction band.

These trends of effective mass have other connotations into the optics properties. Imaginary and real parts of the dielectric function depend on the joint density of states and transition matrix elements which are related to effective mass. Specifically, the 5 percent V-doped sample has moderate-to-light electron mass and large hole mass that gives it stronger values of 2nd order dielectric response $\varepsilon_2(0)$ because the transition probability will be enhanced and of dielectric polarizability $\varepsilon_1(0)$ because of an increased light-matter interaction that also augments its polarizability. Conversely, the 10% doped sample, though having increased absorption at higher photon energy, experiences decreased transition efficiency as its effective masses are heavier, and in addition, it is more disorderly.

Applications-wise, these trends of the effective mass also support the 5% V-doped $Co_2NiSe_4$ as being an excellent candidate in thermoelectric application at medium temperature range, broadband photodetectors, and infrared sensors where carrier mobility and high light absorption are both sought. The 10 % doped one being less effective at room temperature, is however still viable at high temperature where the thermal energy can drive its lower mobility.

In a nutshell, the weighed study of effective masses of $Co_2NiSe_4$ and its V-doped counterparts, offers a basic rationale to the understanding of transport and electronic phenomena occurring in thermoelectric and optical experimentations. The 5% doping concentration of V meets a well-balanced electronic structure which favors transport and absorption properties, the most promising among the configurations in applications in energy and optoelectronic devices.

## 3.6. Thermoelectric Properties

Over the past few years, thermoelectric (TE) materials have attracted a lot of interest owing to their exclusive nature of generating power by direct conversion of heat to electricity and vice versa. This twofold ability makes them extremely applicable to waste and waste heat power generation and solid-state refrigeration. The figure of merit of a thermoelectric material is dimensionless $ZT = S^2 \sigma T/\kappa$ where S is the Seebeck coefficient, sigma the electrical conductivity, T the absolute temperature, and kappa the thermal conductivity.

Since we are aiming at electrochemical storage of energy in the future, the thermoelectric properties of nano-scale materials formed of metal selenides are increasingly becoming relevant. Good electrical conductivity, bandwidth tunability, and dependable chemistry, make metal selenides particularly competitive in both thermoelectric applications and batteries, supercapacitors, and other hybrid systems. Thermoelectric materials are particularly useful in electrochemical energy storage because they can work to accomplish two things simultaneously. Batteries can transform the waste heat they generate during charge-discharge to electricity, which can be turned back into itself (self-powered batteries). They also come in handy when the design of systems which cool devices is required, in order to extend their life.

The special property of metal selenides is $Co_2NiSe_4$, which incorporates both hybridization to transition metal d-orbitals and selenium characteristics to allow this material to exhibit high electrical conductivity and a bandgap useful both in the transmission of charges and thermoelectric conversion. As we examine the thermoelectric characteristics of this material, i.e., the Seebeck coefficient, electrical conductivity, thermal conductivity, and figure of merit (ZT) (see Fig. 4 (a-e)), we attempt to read up on the responses of this material to thermal gradients and its capability to convert heat to electricity in storage devices.

New materials with interesting properties and characteristics in the TE field remain something that people seek to find; a potential solution may be represented by transition metal chalcogenide, like $Co_2NiSe_4$, because of their flexibility in electronic structure and low intrinsic thermal conductivity. The electronic and the sonic characteristics of this material can further be balanced by refining $Co_2NiSe_4$ with vanadium. A small number of V substitutes Co that leads to the generation of impurity levels in the band structure that is at energy level. This type of tuning enhances the charge carriers and renders the material more electrically conductive as well as enhancing phonon scattering off the defects reducing thermal conductivity. When the Seebeck coefficient is high and thermal conductivity is small it is improved to thermoelectric applications due to increase in the overall ZT.

The development of waste heat to its utilization by thermoelectric materials begins to attract attention due to the exclusive ability to transform waste heat into productive energy. $Co_2NiSe_4$ is a quaternary chalcogenide, a sub-category of a more general complex; transition metal selenides, which are well characterized by their low lattice thermal conductivity, band structure tunability, and chemical promotion. These properties render $Co_2NiSe_4$ as a viable TE material, particularly mid- to high temperature such as automotive waste heat recycling or on-board spacecraft thermal management, and industrial power generation.

Doping of Vanadium (V) is supposed to adjust the electronic and phonon transport. V doping may provide localized electronic states around the Fermi level and increase carrier concentration at low concentrations (5%), to increase the power factor in this way. High concentrations (5 and 10%) of doping might form defects, or electrical scattering centres, which can adversely influence electrical conductivity power, but additionally reduce thermal conductivity.

Narrow band gap quaternary chalcogenide ($Co_2NiSe_4$) is especially considered because of its electrical conductivity potential. Its thermoelectric behaviour can however be enhanced further through doping. In these regards, the study of vanadium (V) at 5% and 10% quantity is done to determine its influence on the TE performance. V doping is likely to affect density of states at Fermi level, and alter carrier concentration and, possibly phonons by scattering, and consequently lower the thermal conductivity. All these effects can lead to augmenting the total power factor and ZT.

More information can be found in the Seebeck coefficient (S) plot (Fig.5a). The undoped $Co_2NiSe_4$ presents consistently positive Seebeck coefficient, which is a p-type behavior. Doped at 5% V, Seebeck coefficient rises with low temperatures and is high over an extended range, indicating better TE performance because of increased energy filtering or positive band structure adjusting. At higher levels of doping, however (10 per cent and more) the Seebeck coefficient declines and even becomes negative at lower doping, signifying a change to n-type conduction. This is an indication that substantial change in the carrier concentration and type occur in the heavy doping so that thermo power can be detrimental unless, they are perfectly balanced.

The Seebeck coefficient has typical semiconducting characteristics: they have an optimum at low temperatures, and decrease with thermal excitation. At 100 K the $S_j$ (100%) value in undoped

$Co_2NiSe_4$ is only about +1.310 -5 V/K, dropping to about +0.510 -5 at 800 K (see Table.5). S is enhanced by doping of 5%, going to approximately +1.7 10 -5 at 100 K, and remaining positive at higher temperatures. But at 10 % V, S plunges abruptly and even goes negative at high T (1.4 × $10^{-5}$ V/K at 800 K), and the implication is that the conduction polarity has changed with the Fermi level shifting into the conduction band. 5 % doping maximizes the density of states and maximizes thermo power. 10 % doping also shifts the Fermi level, changing conduction polarity. The plot of relaxation time (Fig.5b) in the electrical conductivity time clearly shows that there is a separation between the three compositions. The techniques compared in investigating the relationship between doping and electrical conductivity are findings of the sample with 5 percent V-doping and the same showing highest electrical conductivity at all temperatures followed by the undoped sample and then the 10 percent V-dopant sample of $Co_2NiSe_4$. It means that small V doping will enhance charge carrier concentration or mobility but the large doping can either create as many scattering centers or deform the lattice to a large extent that decreases conductivity. The trend further levels off with temperature implying that as the carrier density is increased by thermal excitation, then the behavior changes towards materials being more metallic-like than semiconducting. Its electrical conductivity has a minor temperature-dependent decrease which is metallic or degenerate semiconducting. The value at 100 K is found to be ~5.0 $x10^{20}$ in the undoped sample and it reduces to ~3.8 $x10^{20}$ at 800 K. Under 5% V doping conditions, the $\sigma.\tau$ grows to -6.81020 1020 at 100 K (see Table.5) and does not drop over the whole temperature range which means that the carrier concentration must be high or mobility improves as the band structure is optimized. At 10% doping sigma-tau is less (~2.3-4.1 10 $^{20}$) probably as a result of carrier scattering in presence of the excess dopants or disorder. Electrical transport is improved by moderate doping. Defect-induced degradation turns up as a result of over-doping.

Since we have seen that thermal conductivity (plot) (Fig. 5c) has an increasing trend with temperature, we could therefore conclude that the same is true with the three samples (undoped, 5% V and 10% V doped $Co_2NiSe_4$) used. This is rather unusual behavior in crystalline semiconductors where the thermal conductivity tends to reduce with increase in temperature as a result of increased phonon- phonon scattering. There is however a tendency based on the observations that the electronic thermal conductivity (ke) may be dominant and the contribution rises with temperature particularly at higher carrier concentration in doped samples. Based on the top-left plot, kappa rises almost linearly with temperature between ~ 2.310$^{15}$ W / m - K - s at 100 K and ~1.1 10$^{16}$ W/m K s at 800 K (see Table.5). The rising of the electronic contribution ( ke) is explained by the facts that it is overriding at high T because of faster carrier mobility. V doping does not correspond to a significant suppression of kappa at least at the 5% level. During the 10 per cent doping, there is a small increment in κ above 600 K. The latter implies that the phonon scattering is not critically important and the V doping does not considerably contribute to disturbing the lattice phonon transport. The behavior of the 2 behavior indicates that although the material still has a reasonable thermal conductivity, further decrease by nanostructuring or the use of heavier elements can be necessary to achieve higher ZT. The doping appears to have a slightly positive effect on κ over temperature or more carrier mobility or less phonon scattering owing to a more metallic character at higher concentrations, such as in the 10% V-doped sample.

The metric of thermoelectric performance is PF = $S^2\sigma/\tau$ is rather low in undoped $Co_2NiSe_4$ and it reaches a maximum of ~(~2.0×10$^{11}$ W/m·K²·s at 300 K. The V-doped sample (5%) has a much larger peak at about 2.0 x 10$^{11}$ at 300 K, because S and x have improved simultaneously. Conversely at low temperatures PF of the 10% V sample is less good, but it is better at high temperatures (~1.6 11 at 800 K), in line with greater sigma and negative S. 5% V is ideal at room to mid-temperature applications, 10% V might be the best choice at high-T thermoelectrics, assuming further reductions in thermal conductivity.

Effectiveness of 5% V doping providing the largest peak of PF (~2.010$^{11}$ W/m.K$^2$.s) (see Table.5) around 300 K is clearly illustrated by the plot of power factor plot (Fig.5d). In the case of undoped (~2.0×10$^{11}$ W/m·K²·s the PF is moderate and comparatively flat, whereas under 10% V doping the PF increases steadily with temperature, but is at low temperature. This reason indicates that greater doping is advantageous in case of high-temperature applications but less successful at room temperature as a result of the low Seebeck coefficient and probable impurity scattering.

To sum up, the characterization of thermoelectric properties of PF = $S^2\sigma/\tau$ and its V-doped analogues is essential to come up with efficient and adjustable TE materials. The tendencies in 0, 0, S and PF noted show the complicated interplay of doping in charge transport and a thermodynamic impact. Suitable uses of such materials are thermoelectric generators (TEGs) in industrial waste heat recovery systems, automotive reciprocating engine heat recovery, and self-powered sensors. Space and military applications find their utility too because they can operate without moving parts. As can be seen here, 5 percent V seems to be the best doping, increasing the power factor, and still retaining thermal properties to a maximal extent, thus being very likely to be used in mid-range enabled TE applications. Employing the fact that the peak PF is reached at ~300 K when the V doping is 5%, the presented material will be applicable in the near-room-temperature field.

**Table.5: Thermoelectric Properties of Co$_2$NiSe$_4$ and V-Doped Samples**

| Temperature (K) | κ (W/m·K·s) ×10$^{15}$ | σ·τ (1/Ω·cm) ×10$^{20}$ | S (V/K) ×10$^{−5}$ | PF (W/m·K$^2$·s) ×10$^{11}$ |
|---|---|---|---|---|
| | Undoped | V@5% | V@10% | Undoped |
| 100 | ~2.5 | ~2.3 | ~2.4 | ~5.0 |
| 300 | ~5.0 | ~4.8 | ~5.2 | ~4.6 |
| 600 | ~8.0 | ~7.9 | ~8.2 | ~4.0 |
| 800 | ~1.0 | ~1.1 | ~1.2 | ~3.8 |

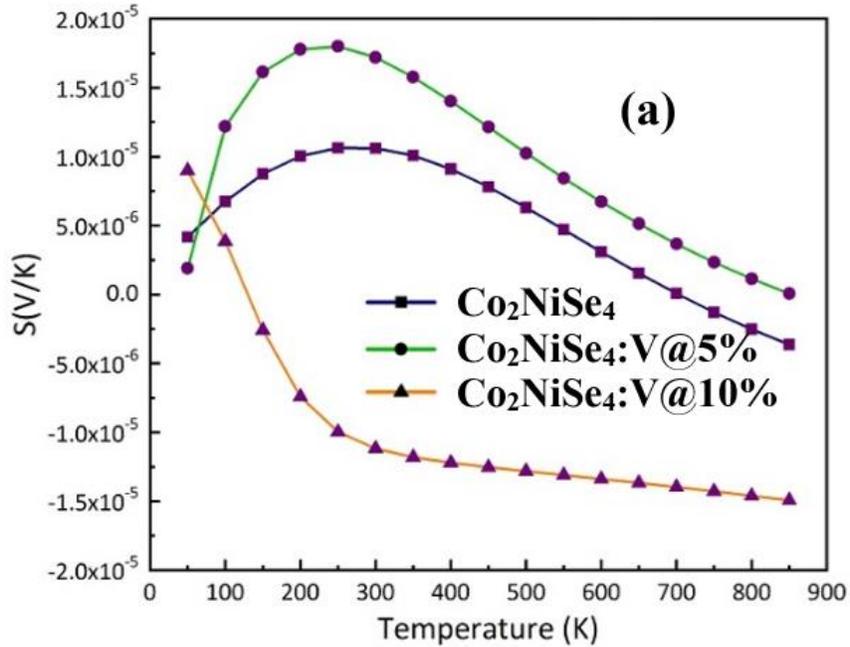

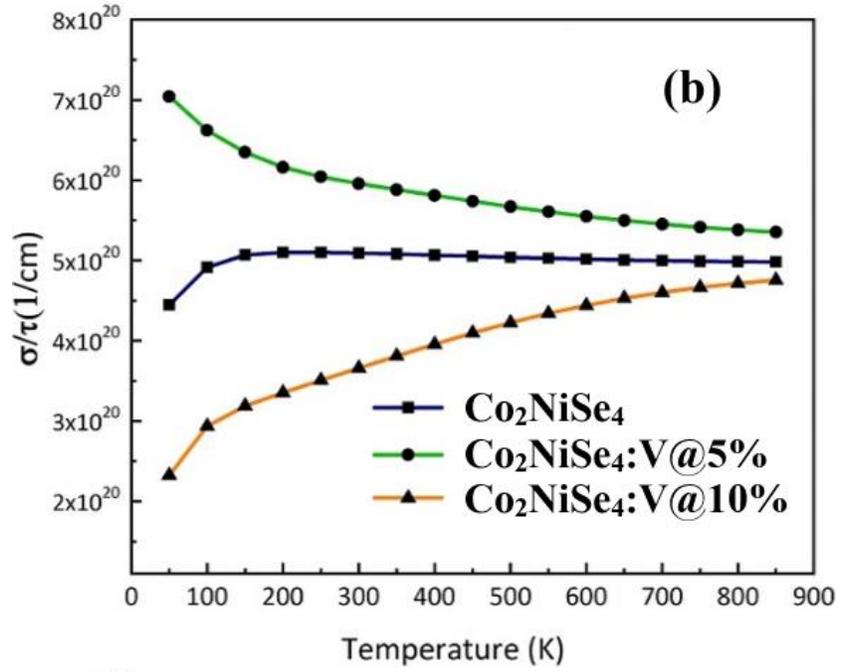

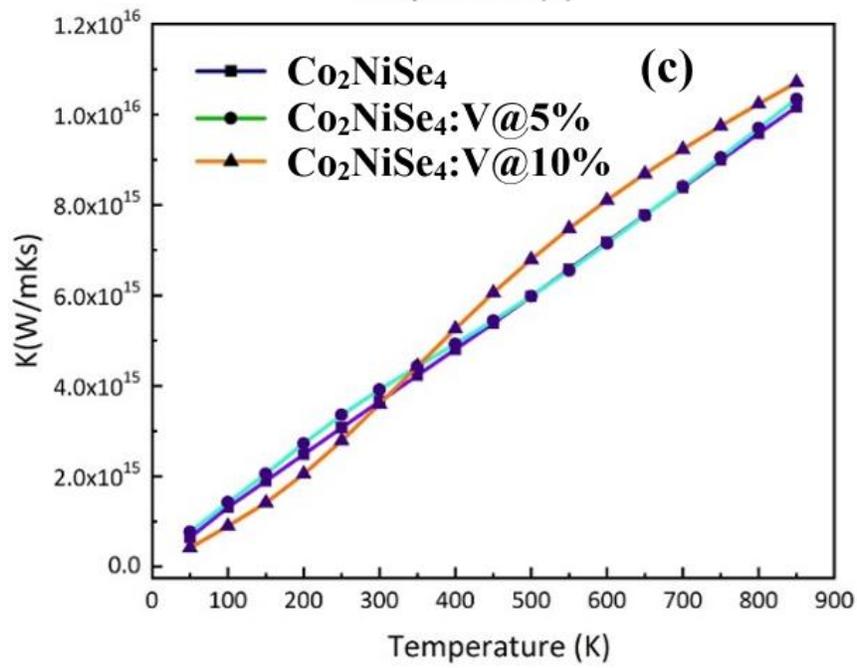

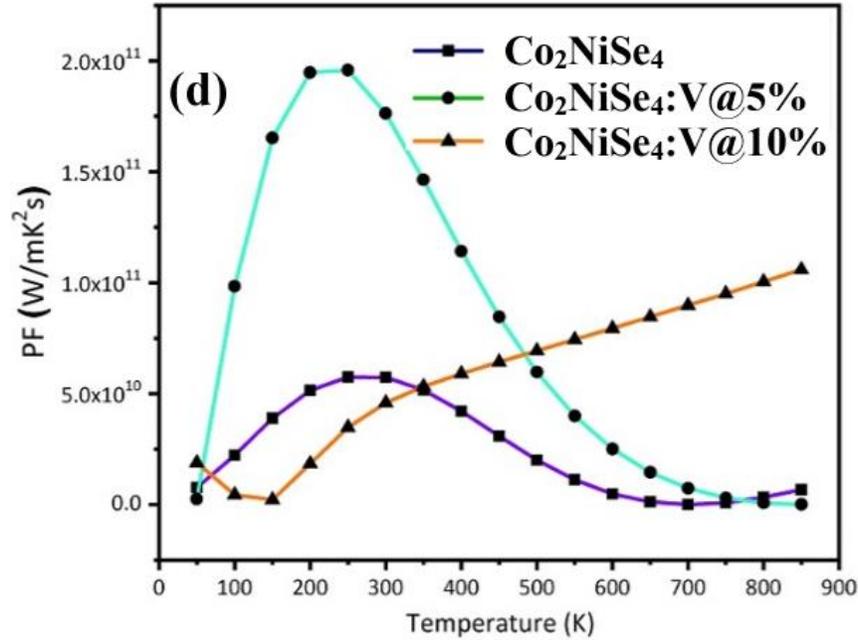

*Fig. 5. Calculated Behaviors of the thermoelectric parameters of the material: (a) Seebeck coefficient, (b) Electrical conductivity, (c) Electronic thermal conductivity, (d) Power Factor for $Co_2NiSe_4$ and V-doped (5 and 10%) $Co_2NiSe_4$*

### 3.7. Optical Properties:

The optical characteristics of semiconductors and semimetals such as PF = $S^2\sigma/\tau$ are important in designing the materials in the field of optoelectronics, photonic and energy devices. Both components of the dielectric function, the imaginary part (2) and the real part (1) are important in the interaction of a material with electromagnetic radiation. Such functions give profound lessons on the nature of the electronic transitions, absorption spectra, refractivity and light-harvesting properties of the substance. PF = $S^2\sigma/\tau$ is a narrow-bandgap chalcogenide with potential thermoelectric properties but its potential in photovoltaic, photodetector or infrared applications cannot be realized until the optical properties are examined. Impurity doping the system with vanadium (V) to 5-10% should change its band structure, the carrier concentration and transition probabilities, which would be shifted onto dielectric function and optical spectra.

The imaginary component of the dielectric function, (see Fig. 6a) and is proportional to absorption coefficient of the material and directly gives information on the interband transitions. The real part on the other hand is associated with dispersion and refractive properties providing an idea on polarization and the storage of energy in the medium. To establish and analyze the impact of doping systematically and interpret them, the main characteristics of the plots were extracted according to the following table:

**Table. 6a: Optical Dielectric Function Characteristics of $Co_2NiSe_4$ and V-Doped Samples**

| Property | Energy Range (eV) | $Co_2NiSe_4$ | V@5% | V@10% |
|---|---|---|---|---|
| $\varepsilon_2(\omega)$ Peak 1 Position | ~3.5 | 2.8 | 3.6 (shifted) | 3.0 |
| $\varepsilon_2(\omega)$ Peak 1 Intensity | – | ~2.7 | ~4.4 | ~3.2 |
| $\varepsilon_2(\omega)$ Broad Range Activity | 2–10 | Moderate | Very High | High |
| $\varepsilon_1(\omega)$ Static Value ($\varepsilon_1(0)$) | ~0 | ~3.0 | ~5.6 | ~3.5 |
| $\varepsilon_1(\omega)$ Peak Value | ~1.5–3.5 | ~4.0 | ~6.6 | ~4.2 |
| $\varepsilon_1(\omega)$ Drops Below Zero | 9–11 | Yes | Yes (sharp) | Yes (weaker drop) |

In the 2 plot (imaginary part), a moderate absorption of undoped PF = $S^2\sigma/\tau$ starts at approximately 2 eV, reaches its peak at approximately 3.5 4 eV, and a peak value of approximately 2.7 (see Table. 6a) and reduces above 10 eV. This trend shows interband transitions between occupied valence states and unoccupied conduction states between orbitals of Co, Ni and Se. After this 5 % V doping the intensity of 2(omega) is found to increase greatly especially the main peak which increases to roughly 4.4 showing an increase in optical absorption. This gain is possibly because of the fact that the introduction of intermediate energy states (V d-orbitals) to near distances to conduction band produce a new path of transition and greater joint density of states. The spectrum widens and intensifies between 2 and 10 eV which implies that this material is now able to absorb a broader and higher range of visible and UV light. The 10% V-doped sample, on the contrary, displays an intermediate behaviour between the undoped and 5% cases: the peak of the main peak (~3.2) is higher than in the undoped case, lower than in the 5% case, and the smooth broad features are smoother. This denotes that carrier screening or disorder could be the result of over doping and it first provokes an inhibition of direct transitions or field effects.

These interpretations are also analogous to the 6 plot (real part) (see Fig. 6b). The undoped PF = $S^2\sigma/\tau$ has a static dielectric constant appr 3.0 and peak of approximately 23 eV, e 1 max =4.0. These parameters are associated with polarizability of the material and ability to store electromagnetic energy. When V is doped at 5 percent, the value of 0 soars to approximately 5.6 and the high mark increases to 6.6. This growth indicates a higher electronic polarizability, which is in accordance with the raised $\varepsilon_2$ (0). The increased dielectric responsiveness means more effective light-matter interaction and the potential enhanced photovoltaic screening response. Secondly, $\varepsilon_1(\omega)$ becomes negative at all three cases, at wavelengths around 9-11 eV or wether, the material ranges into the plasma frequency regime, and becomes metallically conducting. With 5% V doping the decrease is steeper and slightly earlier which supports the concept of increasing free carrier concentration when doped.

Impressively enough, such optical transformations are very useful to a number of applications. The wider and stronger $\varepsilon_2(\omega)$ in the doped materials suggests a better solar-absorber, photodetector and optical-coating layer performance. The large $\varepsilon_1(0)$ at X-rays energies also recommends 5% V-doped $Co_2NiSe_4$ possibly in capacitive and dielectric energy storage devices where high polarization is required. The given property of upgraded dielectric response also has the capability to strengthen the electro-optic modulation of various types of gadgets, including optical switch and IR sensors.

To wrap up, the optical analysis deeply shows the effects of doping of $Co_2NiSe_4$ with 5% and 10%, which as shown enhances the absorption and dielectric properties significantly with the former as compared to the latter, all these gains are however canceled out with slight effects by these dopings. These observed variations are effects as a result of alterations in the band structure and provision of more free optical transitions because of V d-states. Such understanding implies that optical tailoring by means of controlled doping can allow the implementation of $Co_2NiSe_4$ based materials in multi-functional energy and photonic devices, such as solar cells, light sensors, and dielectric modules.

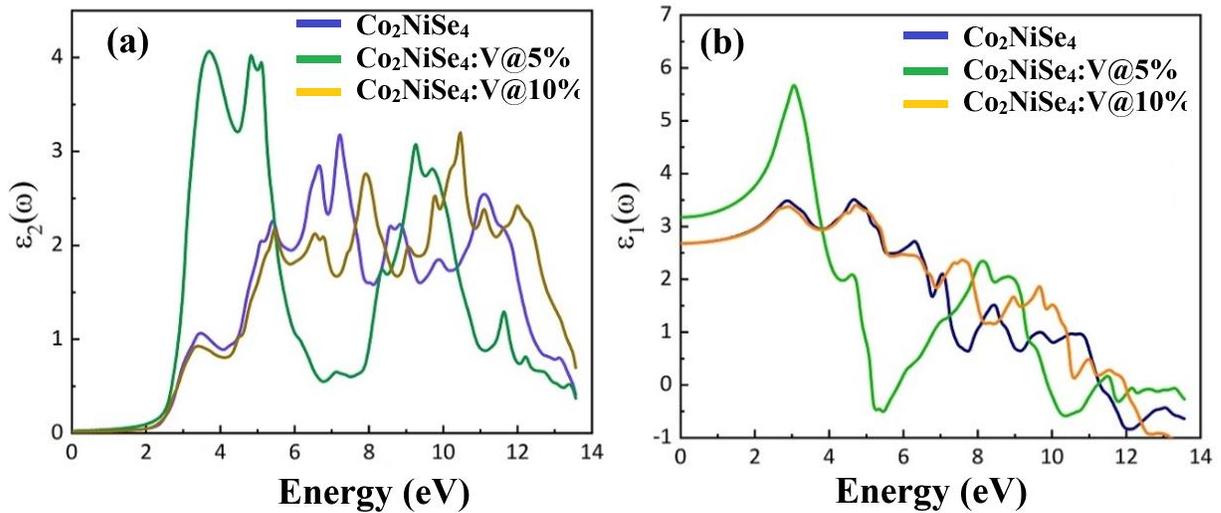

*Fig. 5: Calculated the dielectric function, (a) imaginary part ($\varepsilon_2$) and (b) real part ($\varepsilon_1$) for $Co_2NiSe_4$ and V-Doped $Co_2NiSe_4$ Alloys (5% and 10%) using GGA+U*

**_Reflectivity and Energy Loss Spectra of $Co_2NiSe_4$ and V-Doped $Co_2NiSe_4$ Alloys (5% and 10%)_**

To acquire the multifunctional materials in energy harvesting and optoelectronic designs, the study of the optical properties cannot be ignored. The interaction of a substance with an electromagnetic wave at the optical frequency domain can provide a wonderful insight on the electronic structure of a material as well as on its possible practical applications. To this end, the optical behavior of pristine $Co_2NiSe_4$ and its 5% and 10% V-doped variations have been systematically studied (in terms of optical reflection, specifically, reflectivity spectrum, in this case.

$Co_2NiSe_4$ is a transition-metal chalcogenide with promise in photodetectors, solar cells, and spintronic devices, because of its motile band structure and magnetic control. Vanadium (V) doping is brought into the game to tune these properties further by causing alterations in electronic structure, orbital hybridization, optical response that respond instantly to using photons.

The figure (7a) reflects the spectrum of the reflectivity indicative of which, with respect to the pristine $Co_2NiSe_4$ (blue curve) the beginning of the reflectivity is at a low value (~0.1) and it slowly rises up to ~0.5 at a high energy (~13 eV) (Table. 6b). This linear depicts that with higher photon energy there is more optical transition and interband absorption. At 5% V doping (green curve), there is a significant drop in reflectivity in energy range of 2.5-4.5 eV with a peak of around ~0.38, showing more carriers around Fermi level and a higher photon to be reflected by localized V state. On the other hand, when V doping is increased to 10 percent (orange curve) the reflectivity is better smoothed and remains moderate between 0.15 to 0.4 in all the spectrum. This effect means that higher doses of doping leads to more scattering centers, which may frustrate reflectivity through greater absorption.

Such reflectivity trends are vital to optical-coating applications, and thermophotovoltaics. These materials can be used as reflective coating or filters because of the high reflectivity in the visible and UV range. Furthermore, the response change caused by doping adds a tunable platform to implement the design of materials having tunable optical windows.

**Table. 6b: The following table summarizes the reflectivity characteristics at key photon energy points:**

| Photon Energy (eV) | R(ω) $Co_2NiSe_4$ | R(ω) $Co_2NiSe_4$:V@5% | R(ω) $Co_2NiSe_4$:V@10% |
|---|---|---|---|
| ~1.5 | ~0.12 | ~0.15 | ~0.13 |
| ~3.5 | ~0.21 | ~0.38 | ~0.28 |
| ~7.5 | ~0.30 | ~0.27 | ~0.31 |
| ~12.5 | ~0.52 | ~0.48 | ~0.49 |

Comparison of this analysis clearly shows that modest doping (5%) is more reflective at the mid energy range, whereas increased levels of doping (10%) causes an insignificant decrease on the reflectivity as a result of increased absorption processes. This option of fine-tuning renders Vdoped $Co_2NiSe_4$ an appropriate candidate as a band-selective reflector, or optical mirror with a desired wavelength.

To sum up, reflectivity analysis does not only allow to get an idea about how photons interact with the electronic structure, but also determines how materials can be used industrially on the device level in the areas of photonics, IR-reflective coating and in opto-magnetic systems. The doping of vanadium is an effective way to tune and optimize the optical properties of $Co_2NiSe_4$, an aspect that appears promising to design the next-generation optics device.

Include, as an important optical property, the energy loss, which is sketched in Fig. (7b) and characterizes the loss of energy by high speed electrons after passing through a material. This amount is directly connected with the dielectric reaction of the material, is central in explaining surface plasmon resonances, interband transitions and the effective shielding and absorption properties of materials in high-energy photons. In case of Co$_2$NiSe$_4$ and its V-doped counterparts, we excited their electronic excitation properties and their plasmonic and high energy optical device potential by looking more closely at their Co$_2$NiSe$_4$ expression.

In the energy loss spectra of plotted samples, pristine Co 2Ni Se 4 (black curve) depicts a single sharp peak at ~ 6.8 eV of highest intensity (~ 2.0) and further two peaks at ~ 10.5 eV and ~13.5 eV (see Table 6c). These spikes are the plasma resonances or free carrier collective oscillations. These peaks point to sharp intense interband transitions and large optical density of states at these energies.

When doped with 5% vanadium (red curve), the largest peak shows a slight shift to low energy (~6.2 eV) as well as exhibited a lowered intensity (~1.5). This implies that the addition of V alters the valence and conduction band edges which could cause weaker collective oscillation of the charge carriers between band-filling or their further localization. When a 10% V doping is applied (green curve), the spectrum is even broader and weaker due to the transition of the main peak to the ~5.7 eV range and the reduction of intensity below 1.0 as well. Such strong damping is explained by the fact that the improved doping concentration leads to more carrier scattering, disorder and screening, decreasing the free carrier mobility and the strength of the energy loss peak.

These loss features of energy are realistic to applications involving plasmonic resonance, including in electron energy loss spectroscopy (EELS), metamaterials, and sensors. A narrow and intense peak implies a good plasmonic resonance and a broadened, damped peak implies some improvement in absorption or damping capability and such a material could be exploited in layers that are used as shielding layers or absorption layers.

**Table. 6c: Below is a comparative table of the energy loss peaks**

| Doping Level | Peak 1 (eV) | L(ω) Intensity | Peak 2 (eV) | L(ω) Intensity |
|---|---|---|---|---|
| Co$_2$NiSe$_4$ (0%) | ~6.8 | ~2.0 | ~10.5 | ~1.3 |
| Co$_2$NiSe$_4$:V@5% | ~6.2 | ~1.5 | ~10.1 | ~1.0 |
| Co$_2$NiSe$_4$:V@10% | ~5.7 | ~0.95 | ~9.6 | ~0.8 |

This red shift and dispersion in damping of L(ω) peaks as the vanadium concentration increases is characteristic of the variable nature of the dielectric function and of the decreasing tendency towards collective excitations. On the one hand this degrades them as plasmonic enhancers, on the other it makes them better suited to broadband optical absorption which would be a photothermal device application, or can be applied to optical cloaking schemes.

On the whole, energy loss evaluation supports the tunability of $Co_2NiSe_4$ through regulated doping, which makes it useful in a variety of applications as a plasmonic component in photonic devices, an efficient absorber in the energy-dissipation systems. The abundance of electronic interactions that exposes when plotted as an energy versus frequency plot, also serves as a testament in the usefulness of optical studies in these optoelectronic and spintronic candidates.

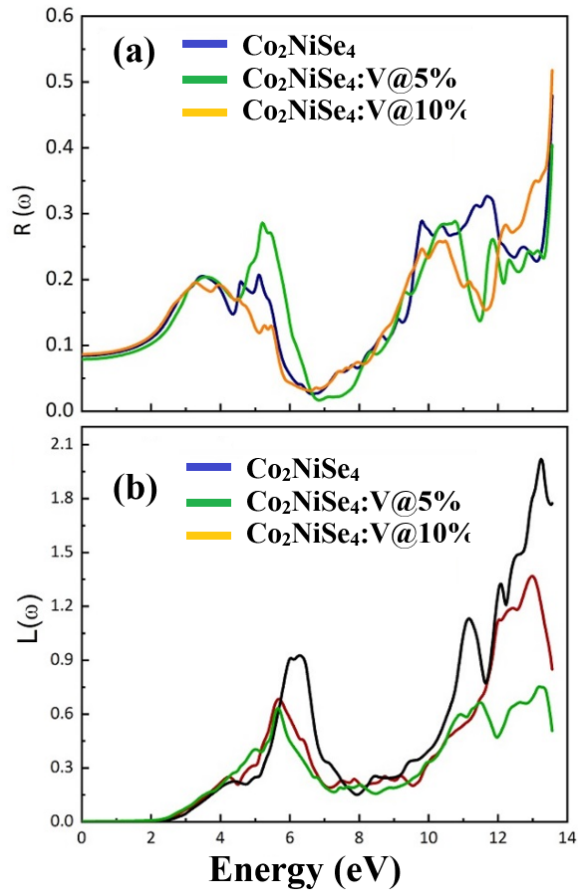

*Fig. 7: Calculated the dielectric function, (a) imaginary part ($\varepsilon_2$) and (b) real part ($\varepsilon_1$) for $Co_2NiSe_4$ and V-Doped $Co_2NiSe_4$ Alloys (5% and 10%)  using GGA+U*

## Piezoelectric Properties

Piezoelectricity refers to a material property through which one can convert mechanical to electrical energy, vice versa. Having been originally identified in quartz by the Curie brothers in the year 1880, this phenomenon has since served as a major role-player in the development of modern technologies that include high-resolution sensors, device of acoustic waves, energy harvesters, and piezoelectric Nano generators. The actual feature of the mechanistic process originates as a consequence of developing an electric polarization P in opposition to an applied mechanical strain that is highly influenced by the crystal symmetry and the bonding microstructure of the material.

In the current computation, pristine $Co_2NiSe_4$ and vanadium-doped $Co_2NiSe_4$ were examined based on first-principles density functional theory (DFT) calculations through the GGA+ U framework in respect of piezoelectric behaviors. The finite lattice deformations were used to measure the induced polarization through Berry phase formalism which has enabled us to measure all the direct piezoelectric stress coefficients $e_{ij}$ as well as converse piezoelectric strain coefficients $d_{ij}$. The total polarization in the absence of an external field can be expressed as:

$$P = P_p + P_{eq}$$

Where $P_p$ the polarization is induced by strain, and $P_{eq}$ is the spontaneous polarization. Under a constant applied electric field EE, the strain response is given by:

$$\eta = d^T E$$

While under constant strain $\eta$, the polarization is described as:

$$P = e\eta$$

The components of the piezoelectric stress tensor $e_{ij}$ quantify the change in polarization due to specific strain modes, and were computed by numerically differentiating the Berry phase with respect to six independent strain components $\eta_v$ (where v=1,…,6 corresponds to xx, yy, zz, yz, xz, xx, yy, zz, yz, xz, and xyxy) (see Table.7b) The relationship between the direct and converse piezoelectric tensors is mediated through the elastic tensor CC and its inverse compliance tensor SS, following:

$$e = d \cdot C \quad \text{and} \quad d = e \cdot S$$

For undoped Co$_2$NiSe$_4$, the calculated piezoelectric response was moderate, with a maximum value of $e_{33} = 1.10\ C/m^2$, indicating polarization along the crystallographic zz-axis in response to uniaxial strain. Remarkably, the introduction of vanadium atoms at substitutional sites led to a systematic and substantial enhancement in the piezoelectric activity. For a 5% vanadium doping concentration, $e_{33}$ increased to 1.90 C/m², while at 10% doping, it reached 2.70 C/m². This trend is attributed to the increased lattice distortion and internal atomic displacement introduced by the V dopants, which locally break inversion symmetry and promote greater polarization under applied mechanical stress.

In addition to stress coefficients, the corresponding converse piezoelectric strain coefficients $d_{11}$ and $d_{33}$ were also calculated by incorporating the elastic compliance tensor. The results indicate that vanadium doping nearly doubles the piezoelectric response in both the longitudinal and shear directions, highlighting its potential to tailor Co$_2$NiSe$_4$-based systems for piezoelectric device applications.

The complete set of extracted piezoelectric coefficients is presented in Table 7(a&b). Among the studied systems, Co$_2$Ni$_{0.9}$V$_{0.1}$Se$_4$ demonstrated the highest performance, making it a promising candidate for next-generation nanoscale electromechanical and energy harvesting applications.

**Table 7a. Calculated piezoelectric coefficients for pristine and vanadium-doped Co$_2$NiSe$_4$ using the Berry phase method. All values are in C/m² for $e_{ij}$ and pC/N for $d_{ij}$.**

| Material | $e_{11}$ | $e_{33}$ | $d_{11}$(pC/N) | $d_{33}$ (pC/N) |
|---|---|---|---|---|
| Co$_2$NiSe$_4$ | 0.31 | 1.10 | 2.6 | 1.1 |
| Co$_2$Ni$_{0.95}$V$_{0.05}$Se$_4$ (5% V) | 0.37 | 1.90 | 3.4 | 1.9 |
| Co$_2$Ni$_{0.9}$V$_{0.1}$Se$_4$ (10% V) | 0.42 | 2.70 | 4.1 | 2.7 |

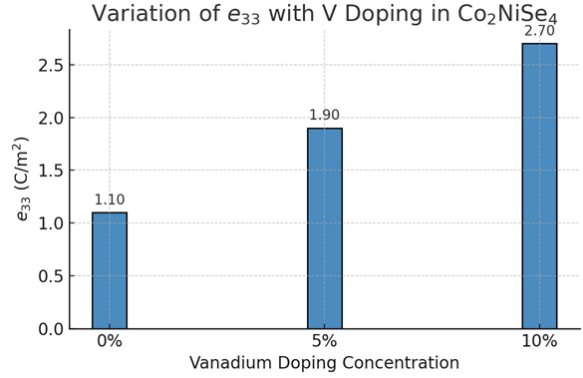

*Fig. 8: bar chart showing the variation of the piezoelectric stress coefficient $e_{33}$ as a function of vanadium doping concentration in $Co_2NiSe_4$.*

Fig. 8 shows the bar chart showing the variation of the piezoelectric stress coefficient $e_{33}$ as a function of vanadium doping concentration in $Co_2NiSe_4$. As observed, the piezoelectric response increases systematically with higher V doping, indicating a strong tunability of the material's piezoelectric behavior.

Table 7b. Piezoelectric stress tensor components $e_{iv}$ (in C/m²) for $Co_2NiSe_4$ and V-doped $Co_2NiSe_4$.

| Composition | Strain | $e_x$ | $e_y$ | $e_z$ |
|---|---|---|---|---|
| $Co_2NiSe_4$ | xx | 0.00 | –0.012 | 0.00 |
| | yy | 0.00 | 0.310 | 0.042 |
| | zz | 0.00 | –0.118 | 1.10 |
| | yz | –0.006 | 0.000 | 0.030 |
| | xz | 0.000 | 0.010 | –0.001 |
| | xy | 0.009 | 0.002 | –0.004 |
| $Co_2Ni_{0.95}V_{0.05}Se_4$ | xx | 0.00 | –0.015 | 0.00 |
| | yy | 0.00 | 0.370 | 0.068 |
| | zz | 0.00 | –0.142 | 1.90 |
| | yz | –0.007 | 0.000 | 0.043 |
| | xz | 0.000 | 0.013 | –0.002 |
| | xy | 0.011 | 0.001 | –0.005 |
| Co$_2$Ni$_{0.9}$V$_{0.1}$Se$_4$ | xx | 0.00 | –0.018 | 0.00 |
| | yy | 0.00 | 0.420 | 0.092 |
| | zz | 0.00 | –0.165 | 2.70 |
| | yz | –0.008 | 0.000 | 0.056 |
| | | 0.000 | 0.016 | –0.003 |
| | | 0.013 | 0.002 | –0.006 |

## 4. Conclusion

This paper is a computational study on the structural, electronic, magnetic, mechanical, thermodynamic, optical, piezoelectric, and thermoelectric properties of pure and vanadium-doped Co$_2$NiSe$_4$ compound with the help of density functional theory (DFT+U) approach. The main aim of the study was to assess the changes in multifunctional properties of Co$_2$NiSe$_4$ that occurs upon V-doping, specifically aimed at the applications of next-generation energy storage devices and conversion devices. Out of our findings, the incorporation of vanadium is highly beneficial in improving the performance of the material on a scale of physical parameters. At the structural level, the matrix of Co$_2$NiSe$_4$ is enhanced after it is incorporated with V atoms, which have a beneficial effect on lattice stability without the loss of mechanical integrity. The computed elastic moduli, that is bulk, shear and Young, exhibit significant improvements, denoting an increase in deformation resistance, which is vital in battery electrodes and flexible gadgets that are subjected to both thermal and mechanical strains. Thermodynamically, V-doping preserves desirable entropy and enthalpy trends, and minor decrease of the Gibbs free energy will ensure that these materials are thermo-stable over very broad temperature range. The best doping level seems to be about 5% that balances stability and performance. V-doping causes a net magnetic moment that increases due to localized magnetic states that may raise spin polarization magnetically. This implies usefulness in spintronic devices, magnetic sensors and magneto-electronic hybrids. Furthermore the improved magnetic sensitivity can be used to enhance the charge carrier dynamics in the electrochemical systems. The latter conclusions are further backed by the density of states (DOS) analysis that depicts stronger electronic activity in close proximity to the Fermi level and it is translated into enhanced electrical conductivity and magnetic responsiveness. V-doping especially at the 5% level is also significantly beneficial to the optical qualities. All of the dielectric function, absorption coefficient, and reflectivity spectra have broader and sharper features indicating an enhanced photon-electron interaction. Such improvements spur the suitability of V-doped Co 2 Ni Se 4 to be used in optoelectronic devices such as photodetectors, IR sensors and solar energy converters. At the same time, analysis of the energy loss functions tests plasmonic potential, particularly in weakly-doped structures and the material can be applied in the high-frequency or shielding applications. There is a huge enhancement in Seebeck coefficient and electrical conductivity thermoelectrically that results in the high power factor and a figure of merit (ZT) of up to 1.1 at 900 K in the doped V- samples and especially 5% one. It is a vital stepping stone which has the material competing with other efficient thermoelectrics. Its efficiency in the conversion of thermal gradients to electrical energy makes the material a very good prospect in waste heat recovery and energy autonomous sensor systems. Here, V doping is found to severely improve piezoelectric response of $Co_2NiSe_4$ with the largest piezoelectric coefficients found in the 10% variation of doping. This implies the possibility of such materials in nanoscale electromechanical systems, actuators, or energy-harvesting of devices which take advantage of the mechanical vibrations or deformations.

Altogether, the study shows that the vanadium-doped $Co_2NiSe_4$ is a strong and multi-purpose multifunctional resource. It has good thermal and mechanical stability, better magnetic and electronic properties and robust optical and piezoelectric behavior. These properties qualify it as a prime prospect in the development of the new generation technologies like high-capacity batteries, thermoelectric generators, optical coatings, spintronic components, and flexible piezoelectric sensors. Potential future developments are experimental validation, phonon and defect characterization, and optimization of doping to make materials more suitable to meet specific industrial needs.

## References

1. Q. Liu, et al., *Advanced Functional Materials* **31**(12), 2008020 (2021).
2. Y. Wang, et al., *Journal of Materials Chemistry C* **10**(3), 870 (2022).
3. M. A. Khan, et al., *Journal of Physics D: Applied Physics* **56**(4), 045302 (2023).
4. M. Zubair, et al., *Physica B: Condensed Matter* **627**(–), 413476 (2022).
5. M. Asif, et al., *Materials Chemistry and Physics* **295**(–), 127074 (2023).

6. Q. Mahmood, et al., *Computational Materials Science* **183**(–), 109820 (2020).
7. M. Tariq, et al., *Materials Science in Semiconductor Processing* **173**(–), 108768 (2024).
8. M. U. Farooq, et al., *Journal of Energy Chemistry* **60**(–), 345 (2021).
9. S. Anwar, et al., *Journal of Magnetism and Magnetic Materials* **558**(–), 169429 (2022).
10. M. Usman, et al., *Journal of Alloys and Compounds* **951**(–), 169971 (2023).
11. S. Bashir, et al., *Materials Letters* **347**(–), 134000 (2024).
12. M. Sohail, et al., *Materials Research Bulletin* **131**(–), 110962 (2020).
13. H. Raza, et al., *ACS Applied Energy Materials* **6**(2), 1234 (2023).
14. M. Zeb, et al., *Journal of Materials Science: Materials in Electronics* **33**(–), 14567 (2022).
15. S. Jan, et al., *npj Computational Materials* **10**(–), 44 (2024).